\documentclass[11pt]{article}

\usepackage{amsmath,amssymb,mathtools,mathrsfs,cite,graphicx,color}
\usepackage{fullpage}
\usepackage[colorlinks=true,urlcolor=blue]{hyperref}
\usepackage{algorithm,algpseudocode}
\usepackage[normalem]{ulem}

\def \endprf{\hfill {\vrule height6pt width6pt depth0pt}\medskip}

\newcommand{\R}{\mathbb{R}}
\newcommand{\C}{\mathbb{C}}
\newcommand{\Z}{\mathbb{Z}}

\newcommand{\E}{\operatorname{E}}

\renewcommand{\d}[1]{d#1}

\newcommand{\e}{e}
\renewcommand{\j}{j}

\newcommand{\vct}[1]{\boldsymbol{#1}}
\newcommand{\mtx}[1]{\boldsymbol{#1}}

\renewcommand{\>}{\rangle}
\renewcommand{\H}{\mathrm{H}}
\newcommand{\T}{\mathrm{T}}
\newcommand{\pinv}{\dagger}

\newcommand{\Span}{\operatorname{Span}}
\newcommand{\trace}{\operatorname{trace}}
\newcommand{\rank}{\operatorname{rank}}

\newcommand{\set}[1]{\mathcal{#1}}

\newcommand\subsetsim{\mathrel{\substack{
  \textstyle\subset\\[-0.2ex]\textstyle\sim}}}

\newcommand{\linop}[1]{\mathcal{#1}}	

\DeclareMathOperator*{\minimize}{\text{minimize}}

\DeclareMathOperator*{\argmin}{\text{arg~min}}

\newcommand{\va}{\vct{a}}

\newcommand{\ve}{\vct{e}}

\newcommand{\vp}{\vct{p}}

\newcommand{\vv}{\vct{v}}
\newcommand{\vw}{\vct{w}}
\newcommand{\vx}{\vct{x}}
\newcommand{\vy}{\vct{y}}

\newcommand{\valpha}{\vct{\alpha}}
\newcommand{\vbeta}{\vct{\beta}}

\newcommand{\veta}{\vct{\eta}}
\newcommand{\vtheta}{\vct{\theta}}

\newcommand{\vphi}{\vct{\phi}}

\newcommand{\vzero}{\vct{0}}

\newcommand{\mA}{\mtx{A}}
\newcommand{\mB}{\mtx{B}}
\newcommand{\mC}{\mtx{C}}

\newcommand{\mE}{\mtx{E}}

\newcommand{\mG}{\mtx{G}}
\newcommand{\mH}{\mtx{H}}

\newcommand{\mL}{\mtx{L}}

\newcommand{\mP}{\mtx{P}}
\newcommand{\mQ}{\mtx{Q}}
\newcommand{\mR}{\mtx{R}}
\newcommand{\mS}{\mtx{S}}

\newcommand{\mU}{\mtx{U}}
\newcommand{\mV}{\mtx{V}}
\newcommand{\mW}{\mtx{W}}

\newcommand{\mZ}{\mtx{Z}}

\newcommand{\mLambda}{\mtx{\Lambda}}

\newcommand{\mSigma}{\mtx{\Sigma}}

\newcommand{\mPhi}{\mtx{\Phi}}
\newcommand{\mPsi}{\mtx{\Psi}}

\newcommand{\mId}{{\bf I}}

\newcommand{\mRbar}{\underline{\mtx{R}}}

\newcommand{\vsbar}{\underline{\vct{s}}}

\newcommand{\vybar}{\underline{\vct{y}}}

\newcommand{\loK}{\linop{K}}

\newcommand{\loP}{\linop{P}}

\newcommand{\setS}{\set{S}}
\newcommand{\setT}{\set{T}}

\newcommand{\vvartheta}{\vct{\vartheta}}
\newcommand{\vtaubar}{\underline{\vct{\tau}}}

\graphicspath{{./Figs/}}

\title{Slepian Beamforming: Broadband Beamforming using Streaming Least Squares}
\author{Coleman DeLude, Mark A. Davenport, and Justin Romberg\thanks{CD,MAD,and JR are all with the School of Electrical and Computer Engineering at the Georgia Institute of Technology.  Email: \{cdelude3,mdav,jrom@ece\}.gatech.edu. This work was supported by COGNISENSE, one of seven centers in JUMP 2.0, a Semiconductor Research Corporation (SRC) program sponsored by DARPA, and a grant from Lockheed Martin.}}

\begin{document}

\maketitle

\begin{abstract}
   In this paper we revisit the classical problem of estimating a signal as it impinges on a multi-sensor array. We focus on the case where the impinging signal's bandwidth is appreciable and is operating in a broadband regime. Estimating broadband signals, often termed broadband (or wideband) beamforming, is traditionally done through filter and summation, true time delay, or a coupling of the two.  Our proposed method deviates substantially from these paradigms in that it requires no notion of filtering or true time delay. We use blocks of samples taken directly from the sensor outputs to fit a robust Slepian subspace model using a least squares approach.  We then leverage this model to estimate uniformly spaced samples of the impinging signal. Alongside a careful discussion of this model and how to choose its parameters we show how to fit the model to new blocks of samples as they are received, producing a streaming output. We then go on to show how this method naturally extends to adaptive beamforming scenarios, where we leverage signal statistics to attenuate interfering sources. Finally, we discuss how to use our model to estimate from dimensionality reducing measurements. Accompanying these discussions are extensive numerical experiments establishing that our method outperforms existing filter based approaches while being comparable in terms of computational complexity. 
\end{abstract}

\section{Introduction}


This paper introduces a new method for estimating a bandlimited signal $s(t)$ impinging on a multi-element sensor array.  Each element of the array observes a version of $s(t)$ that is delayed in time by an amount that depends on the relative position of the sensor to the array center and the angle of arrival of the signal.  The array element outputs are sampled --- we call the collection of samples across the array at a particular time a \emph{snapshot} --- and the goal is to coherently combine these snapshots to estimate a stream of samples of $s(t)$.

When the signal is narrowband,\footnote{We will give a more quantitative way to distinguish ``narrowband'' and ``broadband'' signals in Section~\ref{sec:slepian}, and we will see there that this distinction also depends on the size of the array aperture.} when its carrier frequency is much larger than its bandwidth, a single snapshot contains information about a single sample on the incoming signal (see Figure~\ref{fig:samples}(a) below).  The array samples can be coherently combined across a single snapshot to form an estimate of one sample; this is achieved by taking the inner product of the array snapshot with a set of weights (e.g.\ the steering vector in conventional narrowband beamforming).  

In the broadband case, each snapshot contains information about multiple samples of the signal and each sample of the signal is captured in multiple snapshots (see Figure~\ref{fig:samples}(c)).  The classical approach in this setting is filter-and-sum: the samples at each sensor are delayed by different amounts using an interpolation filter that acts across multiple snapshots, and the filter outputs are combined to form the samples of the beamformed signal. Hence the snapshot-by-snapshot inner product operation in the narrowband case is replaced with a filter bank. 

Our approach is to divide the array snapshots into batches and then to solve a least-squares problem to estimate the signal in overlapping windows of time.  These sequential least-squares problems can either be solved independently or tied together in a manner where estimates in previous regions are updated as new samples are measured (the latter is optimal, somewhat complicated, yet computationally tractable).  Ultimately, the beamformer is a linear map, represented by a matrix, from array samples to signal samples.  We show how this matrix can be designed to point the beam in a particular direction while suppressing energy from other directions in a manner that is directly analogous to the standard MVDR/LMCV narrowband adaptive beamformer.

This approach based on streaming linear algebra allows us to avoid many of the complications (implementing fractional shifts, spectral bleeding, latency introduced by collecting samples across time so they can be transformed into the frequency domain, etc.) inherent to standard methods for broadband beamforming.  We show that in terms of accuracy this is more effective than filter based approaches at estimating a signal due to a reduction in distortion.  Our approach is comparable to standard methods in terms of computationally complexity while enjoying a substantial boost in performance.

Our streaming linear algebra framework also extends to generalized samples.  The Slepian subspace model we use for beamforming also offers a natural way to perform linear dimensionality reduction of the array samples.  In Section~\ref{sec:dimreduce}, we show how our reconstruction algorithm can be modified to work from array samples that have been coded across space and/or time.  

\subsection{Mathematical Formulation}

\begin{figure}[h]
    \centering
    \includegraphics[width = .5\textwidth]{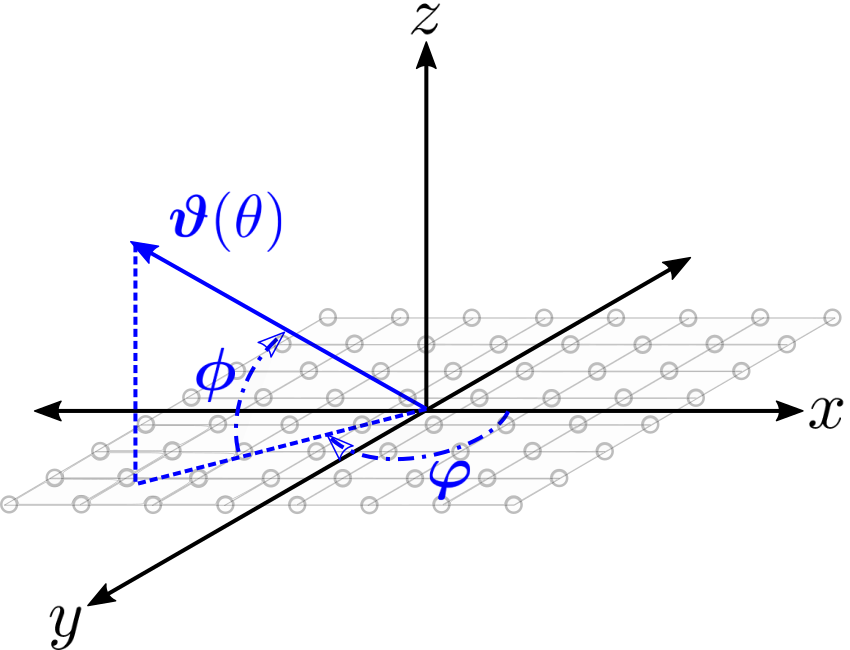}
    \caption{\small \sl In our model the signal $s(t)$ impinges on an array of sensors as a plane wave at a spherical angle $\theta = (\varphi,\phi)$. The sensors are denoted by grey circles and the angular dependent normal vector $\vvartheta(\theta)$ is positioned relative to the phase center of the array. The azimuth $\varphi$ and elevation $\phi$ angles are marked on the figure. Although the above geometry is that of a uniform planar array, these definitions are geometry independent and transfer directly to conformal arrays.}
    \label{fig:arraygeometry}
\end{figure}

Our goal is to estimate a signal $s(t)$ arriving at a sensor array at spherical angle $\theta = (\varphi,\phi)$, where $\varphi$ is the azimuth and $\phi$ is the elevation relative to the array center as shown in Figure~\ref{fig:arraygeometry}.  We assume that the signal has a bandwidth $\Omega$ and has been modulated to carrier frequency $f_c$, so the array observes $s_\mathrm{mod}(t) = s(t)\e^{j2\pi f_c t}$ which lies in the frequency band $[f_c-\Omega,f_c+\Omega]$.

The array has $M$ elements with positions $\vp_1,\ldots,\vp_M$ relative to the array origin.  The framework we develop can be applied to arbitrary arrays in 3D, though of course the effectiveness of the methods we propose will depend on the configuration of the elements inside the aperture (something that is true for any array processing algorithm).  Most of our numerical experiments are for uniform linear and uniform planar arrays with ``$\lambda/2$'' spacing, where $\lambda = c/f_c$ and $c$ is the speed of light.

Adjusting the time axis so that the array center observes $s_\mathrm{mod}(t)$, the modulated signal at element $m$ will be
\[
y_{\mathrm{mod},m}(t) = s_\mathrm{mod}(t-\tau_m(\theta))~+~\text{corruption}, \quad\text{where}~~\tau_m(\theta) = \vp_m^\T\vvartheta(\theta)/c,\quad \vvartheta(\theta) = \begin{bmatrix}
	\cos\varphi\cos\phi \\ \sin\varphi\cos\phi \\ \sin\phi.
\end{bmatrix}.
\]
Demodulating the signal at every element gives us the observational model
\begin{equation}
	\label{eq:ymt}
	y_m(t) = e^{-j2\pi f_ct}y_{\mathrm{mod},m}(t) = \e^{-j2\pi f_c\tau_m(\theta)}s(t-\tau_m(\theta)) ~+~\text{corruption}.
\end{equation}
Thus sampling the demodulated array outputs gives us (noisy) samples of $s(t)$ multiplied by known phases.

We will collect samples off all array elements simultaneously at times $t_1,\ldots,t_N$ ($N$ will vary throughout the paper).  The collection of array output at a fixed sampling time
\begin{equation}
	\label{eq:ysnapshot}
	\vy[n] = \begin{bmatrix} y_1(t_n) \\ \vdots \\ y_M(t_n) \end{bmatrix},
\end{equation}
is called a snapshot.  For exposition, we will assume for most of the paper that the $t_n$ are uniformly spaced and are at (or above) the Nyquist rate for a $\Omega$-bandlimited signal.  The formulation extends easily, however, to situations where the samples are taken at different times at each element and/or are non-equispaced in time.  In Section~\ref{sec:dimreduce} we will also discuss situations where our framework is effective even when the samples are observed through dimensionality reducing measurements.

The collection of snapshots $\{\vy[n]\}_{n=1}^N$ give us information about $s(t)$ over a finite time interval; each sample can be interpreted as a linear measurement of $s(t)$.  We thus treat the estimation of $s(t)$, the beamforming in direction $\theta$, as a linear inverse problem and solve it using least-squares.  The output of the beamformer is a stream of Nyquist samples for the estimated signal.  

We will approach the problem in two different ways.  We will begin in Sections~\ref{sec:conventional} and \ref{sec:adaptive} with static batch estimation where we collect samples over an interval of time and then produce a signal estimate over the same time interval.  In Section~\ref{sec:streaming}, we show how the least-squares solver can be implemented online to produce an estimate of arbitrary length.  Key to both of these approaches is the Slepian representation for bandlimited signals over finite intervals of time which we review in Section~\ref{sec:slepian}. Section~\ref{sec:dimreduce} extends these methods to estimating signals from snapshots that have been linearly encoded across space and time.

\subsection{Narrowband versus broadband}
\label{sec:narrowvbroad}

The observational model described by \eqref{eq:ymt} and \eqref{eq:ysnapshot} can give us a qualitative understanding of what differentiates the narrowband and broadband cases.  Roughly speaking, in the narrowband case the snapshots will be far apart in time (relative to the bandwidth of the signal) with each snapshot essentially containing  information about a single Nyquist sample of $s(t)$.  The array samples in each snapshot can then be coherently combined (usually by phase shifting and summing) independently into estimates of individual samples of $s(t)$.  
In the broadband case, the snapshots overlap with one another; each snapshot contains information about multiple Nyquist samples of $s(t)$ and each Nyquist samples influences multiple snapshots.  Solving the least-squares problem accounts for this in an optimal way.

\begin{figure}
	\centering
	\begin{tabular}{ccc}
		\multicolumn{3}{c}{\includegraphics[width=6in]{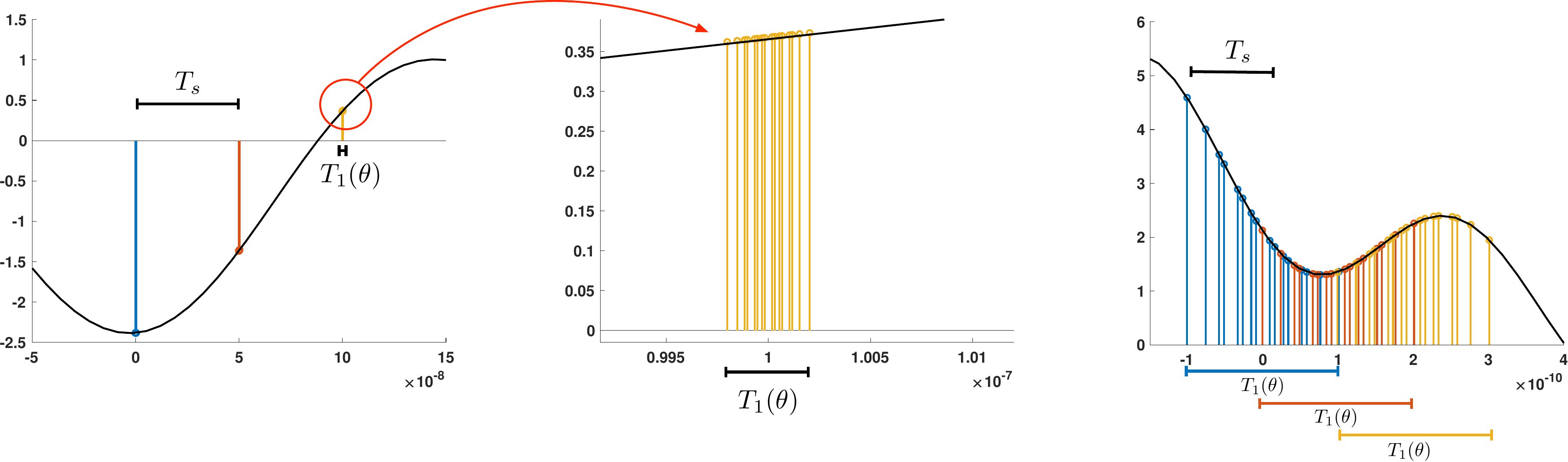}} \\
		(a) narrowband array samples& (b) narrowband zoomed in & ~~~~~(c) broadband array samples
	\end{tabular}
	\caption{\small\sl (a) Array samples of a narrowband signal.  Each snapshot of $M=16$ samples is tightly clustered and can be combined into an estimate of a single sample of the underlying signal.  (b) Zoom in of part (a).  Note that the samples with a cluster are in general non-uniformly spaced.  (c) Array samples of a broadband signal.  Here the snapshots of $M=16$ samples overlap with one another and contain information about multiple Nyquist samples of the underlying signal.  Section~\ref{sec:buls} describes our technique for estimating the signal from these overlapping snapshots.}
	\label{fig:samples}
\end{figure}

These two different situations are illustrated in Figure~\ref{fig:samples}.  In Figure~\ref{fig:samples}(a), we show a narrowband signal being sampled at an array.  The signal has bandwidth $\Omega = 10$ MHz and was modulated to a carrier frequency $f_c = 5$ GHz.  It was incident at angle of incidence $\theta = (\varphi,\phi) = (30^\circ,10^\circ)$ on a $4\times 4$ uniform planar array with spacing $c/2f_c$ matched to the carrier frequency and three snapshots were taken at the Nyquist rate $T_s = 1/2\Omega = 50$ ns.  Each snapshot corresponds to a tight cluster of samples, we zoom in on the cluster for $n=3$ in Figure~\ref{fig:samples}(b).  The length of the time interval that each snapshot covers is 
\begin{equation}
	\label{eq:T1}
	T_1(\theta) =   \max_m\tau_m(\theta)  - \min_m \tau_m(\theta),
\end{equation}
which for this array geometry works out to $T_1(\theta) \approx 0.4$ ns $\ll T_s$.  We are plotting the $\vy[n]$ here without the phases $\e^{-\j2\pi f_c\tau_m(\theta)}$ from \eqref{eq:ymt}, so the signal is close to being a constant over the snapshot time $T_1(\theta)$.  Each cluster can thus be coherently combined into an estimate of a single Nyquist sample of the underlying waveform just as is done in conventional narrowband beamforming.

A broadband example is shown in Figure~\ref{fig:samples}(c).  Here we generated a signal with $\Omega = 5$ GHz, $f_c = 10$ GHz and took $N=3$ snapshot of array samples as in \eqref{eq:ymt}-\eqref{eq:ysnapshot} for a $4\times 4$ uniform planar array with elements spaced at half-wavelength $c/2f_c$m and the same angle of incidence $(30^\circ,10^\circ)$ as above.  In this case, the length of the time interval for a single snapshot $T_1(\theta) = \max_m\tau_m(\theta)  - \min_m \tau_m(\theta) \approx 0.2$ns exceeds the Nyquist sampling interval $T_s = 0.1$ns by about a factor of $2$.  As such, the snapshots, each of which consists of $16$ samples, exhibit significant overlap; the snapshot for $n=1$ is shown in blue, $n=2$ in red, and $n=3$ in yellow. 

The snapshot overlap in the preceding example is actually mild.  For example, one of our numerical experiments in Section~\ref{sec:conventionalexperiments} below uses a $32\times 32$ uniform planar array with $\Omega = 5$ GHz and $f_c=20$ GHz.  With the same angle of arrival as before, we have $T_1(\theta) \approx 1.04$ ns which is a factor of $10$ larger than $T_s = 0.1$ ns, meaning that consecutive snapshot have about $90\%$ overlap.

As we can see from these examples, the quantities $T_s$ and $T_1(\theta)$ give us a natural way to decide if we are in the broadband or narrowband scenarios.  We can safely use narrowband beamforming when $T_1(\theta)\ll T_s$ (so $2\Omega T_1(\theta)\ll 1$) for all angles of arrive $\theta$ of interest; we are plainly in the broadband scenario when $T_1(\theta)\geq T_s$ (so $2\Omega T_1(\theta) \geq 1$) for some $\theta$ of interest.  Setting an exact threshold for $2\Omega T_1(\theta)$ to differentiate the two cases depends on our tolerance for error in approximating the $\{s_m(t-\tau_m(\theta))\}_{m=1}^M$ as a constant.  More precise guidelines are discussed in Section~\ref{sec:slepian} below.

We close this section by reiterating that the notion of broadband in our context does depend on the angle of incidence, especially on the elevation $\phi$.  For $\phi\approx 0$ (meaning the signal is arriving ``broadside''), the samples will always be clustered as in Figure~\ref{fig:samples}(a), and in fact for $\phi=0$ all of the $\tau_m(\theta)$ will be equal to one another and $T_1(\theta) = 0$. The methodology we present below, however, adapts naturally to these cases.


\subsection{Contrast to related work}

The standard beamforming model for both the narrowband and broadband case is concisely over-viewed in \cite{vanveen1988be}. The narrowband case, which this paper does not focus on, takes the form of a weight and sum of sensor outputs. The classically stated condition for the narrowband assumption to hold is that the bandwidth should be small enough for signals received at any point across the array aperture to remain correlated to each other\cite{compton1988ad}. Though related, this a different condition than we formally state in Section~\ref{sec:slepian}. We define the broadband regime to be the point where the subspace dimension required to meet an approximation threshold exceeds one, and the dimension is a function of the signal's time-bandwidth product.

The typical broadband (sometimes termed wideband) beamformer manifests as a set of tapped delay line filters attached to the sensor outputs. A popular assumption is that that the signal has been ``pre-steered" e.g.\ the temporal offsets between sensors for a fixed direction of arrival have been compensated \cite[Chap.\ 6]{vantrees02op}. This is generally achieved through true time-delay or fractional-delay filtering. Post pre-steering the filters are designed to combine channels in a manner that optimally filters noise and nulls interferers. The Frost beamformer \cite{frost1972an} is a popular time-domain implementation of a ``mininum variance distortionless response" (MVDR) beamformer that explicitly leverages pre-steering in its model. It is the time domain extension of Capon beamforming, which in its original formulation is a frequency domain approach \cite{capon1969hi}. As a general statement, essentially every broadband beamformer has complimentary frequency and time domain implementations. However, their respective performance do not substantially differ \cite{compton1988th}. Our methodology distinguishes itself from these by requiring no notion of pre-steering. 

The Frost/MVDR beamformer can be seen as a special case of a broader class of ``linearly constrained minimum variance" (LCMV) beamformers \cite[Chap.\ 2]{liu2010wi}, the main distinction being that the latter in general do not use pre-steering. The fractional delay filtering portion of the optimal filter design is wrapped into the optimzation program through linear constraints. Though this is often done by spectrally sampling the passband, \cite{buckley1987sp} implicitly proposed using Slepian basis vectors to form constraints. Several other works \cite{mayhan1981wi,rodgers1979ad,vook1992ba} examine the performance of tapped-delay line filter based methods with respect to the number of taps. The main focus being how to determine the minumum number of taps required in order to meet certain performance thresholds as a function of the array geometry and bandwidth. Perhaps most similar to our Slepian based MVDR \cite{pezeshki2008ei} explicitly uses Slepian spaces to form a multi-channel output. However, the approach only considers a single snapshot's Slepian space and is mainly focused on robustness to DOA estimation errors. Though we constrain and null our Slepian subspace based methods in a manner similar to these optimal methods, ultimately it is not a filter based approach.

Actually implementing tapped delay lines can be difficult and costly. To circumvent this \cite{Lin2007pe,liu2007ad,liu2009ad} propose ``sensor delay lines" as a method for practically implementing filters. In this paradigm a standard sensor array has a series of additional sensors placed behind it, increasing the array dimension. These additional sensors receive the signal at different times due to the spatial offset, and act as a tap in the standard filtering paradigm. The design process is essentially the same as before, with the main difference being that instead obtaining samples with temporal offsets via a delay line the signal is spatially sampled at the desired offsets. 

Frequency invariant beamforming is a relatively modern approach to broadband beamforming. It is powerful because it allows us to directly map a narrowband beampattern (and all our intuition behind it) to a broadband filter bank \cite{liu2008de,ward1995th}. The method takes a target beam (generally a narrowband beam pattern), and carefully applies a non-linear map that shifts and dilates the target beam frequency response. Mapping this back into the time domain produces a filter bank that will generate the desired beam response across a wide variety of frequencies. A shortcoming of this design methodology is that it attenuates energy at lower frequencies \cite[Chap.\ 5]{liu2010wi}. Hence if the signals energy is dispersed across the entire sampled band the frequency invariant beamformer will induce distortion. Again, our approach is not filter based, and furthermore does not attenuate low frequencies.

Subarray beamforming \cite{nickel2006pr} decomposes the beamforming process into two parts. The operations are chosen such that the first part is generally suitable for analog processing while the second is generally suitable for digitial processing. In the broadband regime the full array is split into subarrays and narrowband beamformed on the subarrays. The ouput of the subarrays is then broadband beamformed \cite{mailoux2009su}. The benefit being the true time delays required to broadband beamform only needs to be applied to the subarray outputs, reducing the number of true time delays the system requires. Performance depends on the size and shape of the subarray, with \cite{nickel1995su,haupt2002op,xiong2013su} examining different partitioning methods. These principles can be extended to optimal beamforming as discussed in \cite{xie2020an}. We note that the basic principle of subarray beamforming is similar to the hybrid beamforming concepts used in MIMO communications \cite{molisch2017hy}, albeit in a more general context. As will be shown explicitly in Section~\ref{sec:subbuls} we can extend our method to the subarray setting, but again we do so without filtering. 

The overarching feature distinguishing our method from all of the described existing approaches is the fact we do not use filters. Fundamentally, filters map blocks of snapshots to a single point whereas our approach maps blocks of snapshots to several points. It does this by successively fitting a robust subspace model to the sample blocks, essentially replacing the application of a filter with vector matrix multiplies. Throughout this paper we provide numerical experiments that compare our method against those mentioned above in a variety of scenarios. Through these results, it is shown that our formulation has the potential to yield substantial performance improvements over existing approaches.

\section{Slepian subspace model}
\label{sec:slepian}

Our approach in Section~\ref{sec:conventional} below treats the estimation of $s(t)$ from the samples taken from multiple array snapshots (as illustrated in Figure~\ref{fig:samples}) as a linear inverse problem.  Like any linear inverse problem, a key part of the solution is an accurate discretization of the signal of interest, which in this case is a bandlimited signal observed over a finite time interval.  There is an excellent, classical way to represent such signals: as a linear combination of a (relatively small number of) prolate spheroidal wave functions \cite{slepian61pr,landau61pr,landau62pr}.  For brevity, we will refer to these as \emph{Slepian functions} and the \emph{Slepian basis}.  Detailed overviews of this representation can be found in \cite{hogan12du,moore04pr}.

The Slepian basis $\{\psi_n(t)\}_{n=1}^\infty$ is an orthonormal decomposition for signals on an interval which we will take as $[0,T]$ for our exposition.  Any $s(t)\in L_2([0,T])$ can be written as
\begin{equation}
	\label{eq:slepianexpansion}
	s(t) = \sum_{n=1}^\infty \alpha_n\psi_n(t),\quad t\in[0,T],
\end{equation}
where $\alpha_n = \<s(t),\psi_n(t)\>$.  There is no closed form expression for the Slepian basis functions, but they are the eigenfunctions of the integral operator $\loK:L_2([0,T])\rightarrow L_2([0,T])$ defined as
\begin{equation}
	\label{eq:slepiankernelop}
	\loK[\vx](t) = \int_0^T k(t,s)x(s)~\d{s},
	\quad k(t,s) = \frac{\sin(\Omega(t-s))}{\pi(t-s)},
\end{equation}
for a bandwidth parameter $\Omega$.  

If the signal $\{s(t),~t\in\R\}$ on the entire real line is bandlimited to $\Omega$, then it can be approximated very 
closely by truncating the sum in \eqref{eq:slepianexpansion} to a finite number of terms.  It is a classical result  \cite{slepian76ba} that truncating to $d\approx 2\Omega T$ terms will provide a good approximation for the vast majority of bandlimited signals.

 We make this precise the in following manner.  Let $x(t)$ be a Gaussian random process that is bandlimited with a flat power spectrum over the frequency band $[-\Omega,\Omega]$ which is observed on the time interval $[0,T]$.   Let $\loP_d[x](t)$ be the projection onto the approximating subspace of dimension $d$,
\[
	\loP_d[x](t) = \sum_{k=1}^d \beta_n\psi_n(t),
\]
where again $\beta_n = \<x(t),\psi_n(t)\>$.  For a fixed $\epsilon > 0$, we define the dimension of $\Omega$-bandlimited signals over $[0,T]$ as
\[
	d(\Omega T) = \argmin_{d\geq 1} \left\{d~:~\E\left[\int_{0}^T|x(t)-\loP_d[x](t)|^2~\d{t}\right]\leq\epsilon\cdot\E\left[\int_{0}^T|x(t)|^2~\d{t}\right]\right\}.
\] 
This basically says that we are choosing $d$ so that a generic bandlimited signal will have at least $(1-\epsilon)$ of its energy in the subspace spanned by the first $d$ Slepian basis functions.  While we omit this it in the notation, it should be understood that this dimension depends on the tolerance parameter $\epsilon$.

There are known upper and lower bounds on $d(\Omega T)$.  Classical asymptotic bounds can be found in \cite{landau80ei,landau93de}, while results in the recent literature provide more precise non-asymptotic bounds \cite{osipov13ce,israel15ei,karnik21im,bonami21no}. For example,  \cite[Cor.\ 4]{karnik21im} shows that we can take
\begin{equation}
	\label{eq:dTtheory}
	d(\Omega T) \leq  \lceil 2\Omega T\rceil + c_1\log(2\Omega T)\log(1/\epsilon) + c_2,
\end{equation}
for $\epsilon < 1/2$.  The $c_1$ and $c_2$ are constants that can be calculated and refined for different ranges of $\Omega T$.

While the theoretical guarantees like \eqref{eq:dTtheory} are informative, we can also get very tight estimates for $d(\Omega T)$ through numerical computations.  Let $\lambda_1\geq\lambda_2\geq \cdots$ be the eigenvalues of the integral operator in \eqref{eq:slepiankernelop} associated with the eigenfunctions (the Slepian basis functions) $\psi_1(t),\psi_2(t),\ldots$.  An equivalent definition of $d(\Omega T)$ can be written in terms of these eigenvalues:
\[
	d(\Omega T) = \argmin_{d\geq 1} \left\{d~:~ \sum_{k=d+1}^\infty\lambda_k \leq \epsilon\sum_{k=1}^\infty\lambda_k = \epsilon\,\Omega T\right\}.
\]
There are known numerical methods to compute the $\{\lambda_k\}$ to arbitrary precision; see for example \cite[Chap.\ 8]{percival93sp}.  Using these methods\footnote{Code for these results is available at \url{https://github.gatech.edu/cdelude3/BroadbandBeamforming.git}} we can compute $d(\Omega T)$ for particular $\epsilon$ and $\Omega T$ of interest.  For example, for $\epsilon = 10^{-3}$ we have
\begin{equation}
	\label{eq:dt10m3}
	\begin{cases} 
		d(\Omega T) = 1, & \Omega T \leq .031, \\ 
		d(\Omega T) = 2, & .032\leq\Omega T \leq 0.268 \\
		d(\Omega T) \leq  \lceil 2\Omega T\rceil + 2, & \Omega T\leq 3.4 \\
		d(\Omega T) \leq \lceil 2\Omega T\rceil + 3, & \Omega T \leq 200,
	\end{cases}
\end{equation}
and also $d(\Omega T)\geq \lceil 2\Omega T\rceil + 1$ for $\Omega T \geq 0.32$.  Similarly, for $\epsilon = 10^{-4}$ we have
\[
\begin{cases} 
	d(\Omega T) = 1, & \Omega T \leq .009, \\ 
	d(\Omega T) = 2, & .010 \leq\Omega T \leq  0.151\\
	d(\Omega T) = 3, & .152 \leq\Omega T \leq  0.439\\
	d(\Omega T) \leq  \lceil 2\Omega T\rceil + 3, & \Omega T\leq  3\\
	d(\Omega T) \leq \lceil 2\Omega T\rceil + 4, & \Omega T \leq 200,
\end{cases}
\]
and also $d(\Omega T)\geq \lceil 2\Omega T\rceil + 3$ for $\Omega T \geq 1$.  These results give us very precise information about the error incurred when the basis expansion sum in \eqref{eq:slepianexpansion} is truncated.

In the sections below, we will use the Slepian basis to model the underlying signal over the time interval encompassed by a series of array snapshots.  We have already encountered, in \eqref{eq:T1} in Section~\ref{sec:narrowvbroad}, the time interval $T_1(\theta)$ that contains the (in general non-uniformly spaced) samples for a single signal array snapshot; this is a function of the array geometry (which in turn might be an implicit function of the carrier frequency $f_c$) and the angle $\theta$.  For a uniform linear array with $M$ elements spaced at half-wavelength spacing of $c/2f_c$, we have
\[
	0\leq T_1(\theta) \leq \frac{M}{f_c},
\]
where the lower bound is met when the signal is arriving ``broadside'' ($\phi=90^\circ$) and the upper bound is met when the signal is ``endfire'' ($\phi = 0^\circ$).  For an $M$-element uniform planar array (UPA) consisting of elements placed on a  $\sqrt{M}\times\sqrt{M}$ grid at half-wavelength spacing, we have
\[
	0\leq T_1(\theta) \leq \frac{\sqrt{2M}}{f_c}.
\]
Note the square-root dependence on the number of elements for the UPA which arises because the effective aperture of the array scales with its diameter.

The bounds on the time interval $T_1(\theta)$ and the corresponding dimension in \eqref{eq:dt10m3} give us direct information about the number of degrees of freedom in an array snapshot.  For the narrowband example illustrated in Figure~\ref{fig:samples} (bandwidth $\Omega=10$ MHz, carrier frequency $5$ GHz, $4\times 4$ array) we have $\Omega T_1(\theta) = .004$ and so $d(\Omega T_1(\theta)) = 1$ for $\epsilon = 10^{-3}$.  Thus the information about $s(t)$ over the interval $T_1(\theta)$ can be captured with a single linear measurement.  For conventional narrowband beamforming, this measurement is $\hat{s}[n] = \va_\theta^\H\vy[n]$, where $\va_\theta = \{e^{j2\pi f_c\tau_m(\theta)}\}_{m=1}^M$ is the steering vector for the array in direction $\theta$.  In Section~\ref{sec:conventionalexperiments} below, we simulate a signal with $\Omega = 5$ GHz, $f_c = 20$ GHz incident on a $32\times 32$ array; in this case we have $d(\Omega T_1(\theta)) = 14$, meaning that it takes roughly $14$ terms in \eqref{eq:slepianexpansion} to capture $s(t)$ over the interval $T_1(\theta)$.

In general, we will be interested in estimating $s(t)$ from batches of array snapshots.  If we take $N$ samples with spacing $T_s = 1/2\Omega$, then the time interval over which we observe the signal is
\[
	T_N(\theta) = T_1(\theta) + \frac{(N-1)}{2\Omega},
\]
meaning that the dimension of our representation is
\begin{equation}
    \label{eq:dN}
    D_N(\theta) := d(\Omega T_N(\theta)) = \lceil 2\Omega T_1(\theta)\rceil + L + N - 1,
\end{equation}
for some $L$ that depends very weakly on the tolerance $\epsilon$ and the number of samples $N$ (the bound in \eqref{eq:dTtheory} suggests that $L$ should grow as $\log(1/\epsilon)$ and $\log N$).  Note that as $N$ grows, $D_N(\theta)/N\rightarrow 1$ so the dimensionality approaches one degree of freedom per Nyquist sample, as expected for a bandlimited signal.

In the next section, we will see how we can use the Slepian subspace model to perform broadband beamforming.

\section{Conventional beamforming}
\label{sec:conventional}

In this section, we show how to perform beamforming from a series of snapshots taken off the array.  By ``beamforming'' we mean forming an estimate of the signal under the hypothesis that it is arriving from a given direction $\theta$ (this is the same language used in the classic text \cite{vantrees02op}).  We treat the process of beamforming as solving a linear inverse problem using least-squares by combining the observational model in \eqref{eq:ymt} (that explicitly depends on the angle $\theta$) with the linear subspace model \eqref{eq:slepianexpansion} provided by the Slepian basis.  We ultimately construct a linear map from array snaphots to Slepian coefficients of the signal.  These coefficients can in turn be mapped to equispaced Nyquist samples using another linear map.

We start in this section by considering the reconstruction problem over a fixed, relatively small time frame.  In Section~\ref{sec:streaming} below, we will show how this procedure can be made fully online through a streaming least-squares solver  that estimates the signal in overlapping blocks and then combines these estimates in an optimal way.  This allows us to form the beam over arbitrarily long time intervals using modest computational resources.

\subsection{Beamforming using least squares}
\label{sec:buls}

Suppose we collect $N$ snapshots off of an $M$ element array.  As discussed above, these samples contain information about the signal in a time interval of length $T_N(\theta)$ (we will simply take this interval to be $[0,T_N(\theta)]$ for the exposition below).  It is clear that the mapping from the signal $s(t)$ to the array samples is linear.  As such, we will treat the estimation of $s(t)$ as a linear inverse problem.  Since
\begin{equation}
	\label{eq:slepiantruncation}
	s(t) \approx \sum_{k=1}^{D_N(\theta)}\alpha_k\psi_k(t),
\end{equation}
where the $\{\psi_k\}$ are the Slepian functions discussed in Section~\ref{sec:slepian}, we will estimate $s(t)$ by estimating the first \eqref{eq:dN} expansion coefficients $\{\alpha_k\}$.  With this discretization of $s(t)$ in place, we now have a ``forward model'' for a sample snapshot \eqref{eq:ysnapshot},
\begin{equation}
	\label{eq:forwardmodel}
	\vy[n] = \mA_n(\theta)\valpha + \text{corruption},
\end{equation}
where $\mA_n(\theta)$ is a $M\times D_N(\theta)$ matrix with entries
\[
A_n(\theta)[m,k] = \e^{-j2\pi f_c\tau_m(\theta)}\psi_k(t_n-\tau_m(\theta)).
\]
Stacking the $N$ snapshots together with their corresponding forward models into $\vybar$ and $\mA(\theta)$,
\begin{equation}
	\label{eq:yAtheta}
	\vybar = \begin{bmatrix}
		\vy[1] \\ \vdots \\ \vy[N]
	\end{bmatrix},
	\quad
	\mA(\theta) = \begin{bmatrix}
		\mA_1(\theta) \\ \vdots \\ \mA_N(\theta)
	\end{bmatrix},
\end{equation}
we can now estimate the Slepian coefficients by solving the least-squares problem
\begin{equation}
	\label{eq:lsconv}
	\minimize_{\valpha\in\C^{D_{N}(\theta)}}~\frac{1}{2}\|\vybar - \mA(\theta)\valpha\|_2^2 + \delta\|\valpha\|_2^2.
\end{equation}
The $\delta>0$ above is a regularization parameter that can be set depending on the structure of the singular values of $\mA(\theta)$; we will see below that $\mA(\theta)$ is in general fairly well conditioned and so we can typically take $\delta = 0$.  With $\delta=0$, the closed form solution to \eqref{eq:lsconv} is
\begin{equation}
	\label{eq:pseudoinv}
	\widehat\valpha = \mA(\theta)^\pinv\vybar = \mV(\theta)\mSigma(\theta)^{-1}\mU(\theta)^\H\vybar,
\end{equation}
where $\mA(\theta) = \mU(\theta)\mSigma(\theta)\mV(\theta)^\H$ is the singular value decomposition of $\mA(\vtheta)$.

Given $\widehat\valpha$, we can reconstruct any set of samples we like using another linear map.  For example, if we want to estimate $s(t)$ at $t_1,\ldots,t_N$, we simply form the $N\times D_N(\theta)$ matrix $\mPsi$ with entries $\Psi[n,d] = \psi_d(t_n)$ and compute $\widehat\vsbar = \mPsi\widehat\valpha$.  Note that the entire mapping from array samples $\vybar$ to samples $\vsbar$ of the beamformed signal can be collapsed into a single matrix, $\vsbar = \mPsi\mA(\theta)^\pinv\vybar =: \mW(\theta)\vybar$.  The rows of $\mW(\theta)$ can be interpreted as ``weights'' that combine array samples across space and time into an estimate of a sample of $s(t)$. This concept will be discussed with great detail in Section~\ref{sec:dimreduce}.

Once we have formalized the forward model \eqref{eq:forwardmodel} and the Slepian discretization \eqref{eq:slepiantruncation}, the least-squares approach to estimation in \eqref{eq:lsconv} is straightforward and natural.  It it also, as we will show with numerical experiments in the next section, extremely effective as it outperforms both classical and recently developed methods for broadband beamforming while being no more expensive computationally.

\subsection{Numerical experiments: Comparison to delay-and-sum}
\label{sec:conventionalexperiments}

As a first demonstration of the effectiveness of our method this we compare its performance to more standard paradigms. We examine two test cases, the first being the $32 \times 32$ element UPA described in Section~\ref{sec:slepian} where $f_c = 20$ GHz, $\Omega = 5$ GHz, and $\theta = (\pi/4,\pi/2)$. The second case is a 64-element uniform linear array (ULA) where $f_c = 20$ GHz, $\Omega = 5$ GHz, and $\theta = \pi/2$. For the ULA $d(\Omega T_1(\theta)) = 20$ meaning its spatial subspace dimension is slightly larger. In both cases we examine $N=2^5$ snapshots. To test the performance of various beamforming methods on these arrays we generate a random bandlimited signal using a sum-of-sinusoids model. The signal is then corrupted by complex Gaussian noise at a nominal SNR and impinges on the array as a plane wave. The signal is beamformed using one of the described methods and we observe the beamformed SNR. Results were averaged across 50 randomly generated trials.

In terms of conventional methods we compare against two versions of delay and sum beamforming. The first is the more traditional of the two and employs fractional delay filters to digitally align the signals in time (often termed pre-steering) prior to summing \cite{vantrees02op}. These fractional delay filters consist of sinc interpolators truncated to $R$ taps, where we note that as $R\to\infty$ the filters become ideal. Since performance varies drastically with the length of the filter we compare against several choices in $R$. We note that there are several methods of fractional delay interpolation that may offer better results than sinc interpolation. However, methods such a Lagrange or spline interpolation need some degree of oversampling to yield truly drastic improvement over sinc inteprolation. Since we are operating in a critically sampled regime, the difference amongst interpolation methods is limited and therefore we settle for the most intuitive method. The second version of delay and sum we examine is a frequency invariant beamformer, which uses a nonlinear map to generate a broadband filter bank that matches a narrowband response across a desired frequency band \cite{liu2008de,ward1995th}. In this case we map a Hann windowed narrowband response to the frequency invariant broadband regime.

\begin{figure}[h]
	\centering
	\begin{tabular}{cc}
		\includegraphics[width=.45\textwidth]{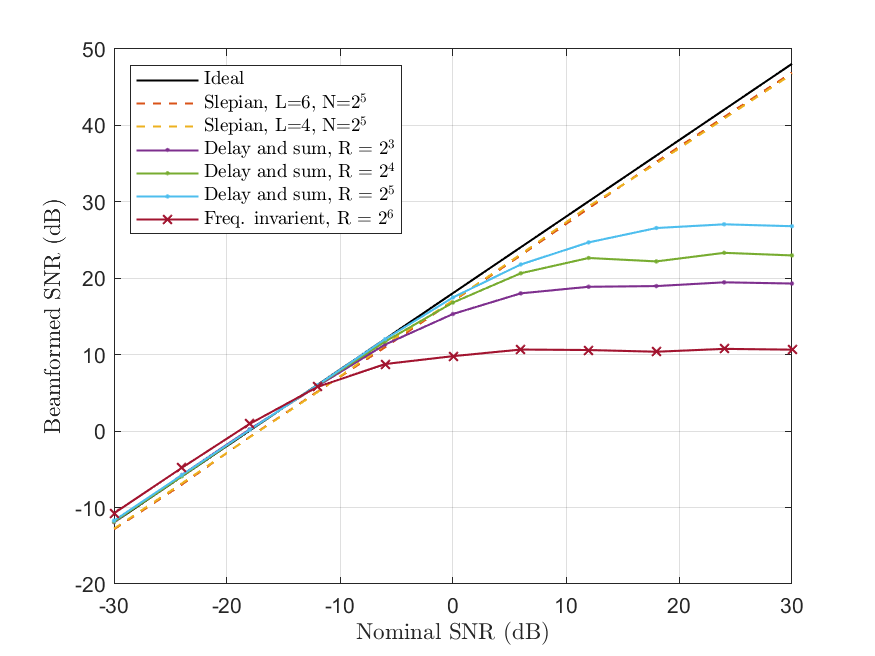}  & \includegraphics[width=.45\textwidth]{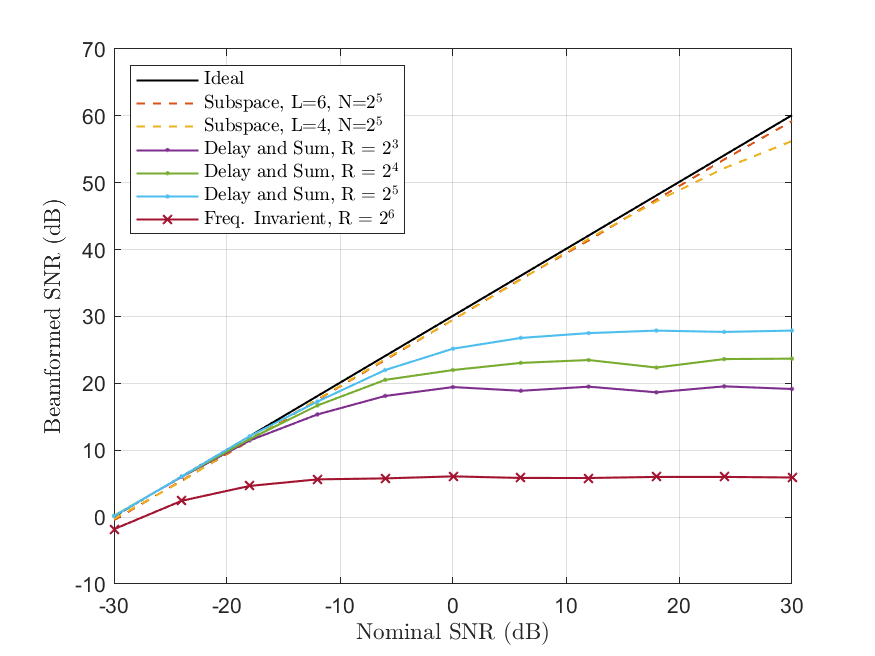}\\
		(a) 64 element ULA & (b) $32\times32$ element UPA
	\end{tabular}
	\caption{\small\sl A comparison of delay and sum beamforming via fractional delay filtering, frequency invariant beamforming, and Slepian subspace beamforming for a (a) 64 element ULA and a (b) $32\times32$ element UPA. A series of nominal SNRs are compared to the subsequent beamformed SNRs, where the boost in SNR is the array gain. The ideal array gain, which can be achieved with perfect delays (e.g.,\ infinite length filters), is given by the black line.}
	\label{fig:conv_bf_ag}
\end{figure}

The results for the ULA and UPA are shown in Figure~\ref{fig:conv_bf_ag}(a) and \ref{fig:conv_bf_ag}(b) respectively, where the black line represents the ideal array gain of $M$. We note that this ideal gain can be achieved when the fractional delay filters are infinite in length, or a perfect true time delay can be applied. The results for both arrays follows a similar trend; at low SNRs the error is noise dominated and hence all methodologies perform similarly. The story changes as we increase the SNR. The error becomes dominated by the filter bias and we see a sharp performance roll off. Though increasing the filter size alleviates this effect to some extent, even for long filters the roll off is apparent. 

Perhaps surprisingly, the frequency invariant beamformers yield comparatively poor performance. This is likely due to attenuation of lower frequency components, which is a known phenomena and discussed in detail in Appendix~\ref{apx:fibf}. Essentially the filter design process has a limited ability to produce filters with good frequency invariance at low frequencies\cite{liu2008de,liu2010wi}. Since we assume baseband processing, critical sampling, and the signal is evenly distributed across the band a substantial portion of the signals energy gets attentuated, leading to high distortion. However, if the signal was oversampled or its energy was concentrated only in the upper half of the band, then this style of beamforming would perform far better. A demonstration of the affects of oversampling on frequency invariant beamforming are provided in Appendix~\ref{apx:fibf}.

In contrast to delay and sum methods, our proposed Slepian subspace beamformers track the ideal array gain across all tested SNRs. The bias induced by the subspace truncation error is drastically smaller than the bias induced by the filter distortion, and furthermore consistent within the tested regime. In Figure~\ref{fig:conv_bf_ag}(b) we can see a small performance roll-off beginning as the ideal beamformed SNR approaches the subspace truncation error. However, this is easily remedied by a modest increase in the $L$ parameter.


\subsection{Subarray beamforming using least squares}
\label{sec:subbuls}
Using our least squares approach we can formulate a natural extension to subarray beamforming. We begin this discussion with a brief introduction into the general concept of subarray beamforming. In the standard case the array is divided into multiple smaller arrays i.e., subarrays. More formally we break an $M$ element array into $M'$ subarrays. The size of the subarrays is chosen to ensure the aperture is reduced to the point that the signal is narrowband as it lies across any particular subarray \cite{nickel1995su,nickel2006pr}. Let $\mathcal{M}_{m'} \subset \{ 1,\dots,M \}$ be the set containing the indices associated with the $m'$th subarray. The signal is then narrowband beamformed with weights $\{w_m\}_{m=1}^M$ at the subarray level such that that a given sub-beam is
\begin{align}
    \label{eq:ysub}
    y_{\mathrm{sub},m'}(t) = \sum_{m\in {M}_{m'}} w_{m}y_m(t).
\end{align}
The temporal offset between sub-beams is traditionally corrected via filtering (or any suitable method) followed by summation to form the full beam. Some advantages of this is that it reduces the number of channels that must be filtered, and it splits the problem into operations that can naturally and respectively be handled in the analog and digital domain \cite{molisch2017hy}.

Given access to sub-beams instead of direct samples off the array changes our forward model such that \eqref{eq:lsconv} must be modified. The details of this modification are thoroughly overviewed in Section~\ref{sec:dimreduce}, but in general as long as $M'>d(\Omega T_1(\theta))$ so that the spatial subspace dimension is sufficiently sampled, we can recover $\valpha$ as accurately as if we had direct access to the $\vybar$.

In the next section we display numerical experiments demonstrating that subarray Slepian subspace beamforming performs nearly as well as the full scaled version. Again our least squares approach outperforms conventional methods, and additionally shows less degradation in performance with the subarray size. The concepts put forth in this section, namely dimensionality reduction prior to coefficient estimation, will be generalized in Section~\ref{sec:dimreduce} where subarray beamforming can be considered as a special case.

To see how subarray beamforming effects performance we return to the test cases presented in Section~\ref{sec:conventionalexperiments}, with all parameters for the ULA and UPA remaining the same. We test two subarray partitionings for each geomery, with the ULA being split into $2 \times 1$ and $4 \times 1$ subarrays and the UPA being split into $2 \times 2$ and $4 \times 4$ subarrays. We again compare the Slepian subspace beamformer's performance to delay and sum beamformers using fractional delay filters. For delay and sum, the subarrays are beamformed using standard complex phase weightings. Time delays are applied to the sub-beams using length $R = 2^6$ sinc-interpolation filters. 

\begin{figure}[h]
	\centering
	\begin{tabular}{cc}
		\includegraphics[width=.45\textwidth]{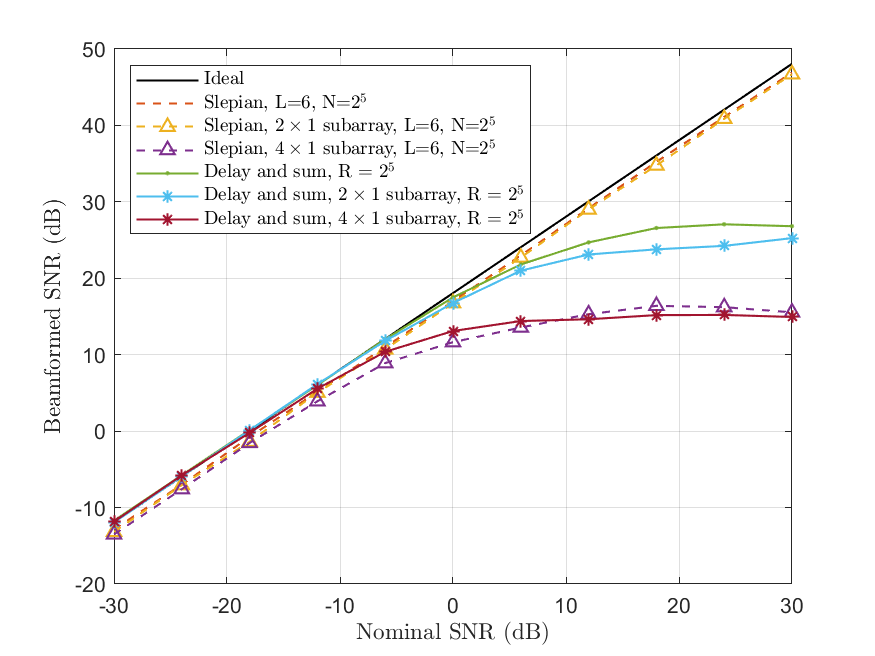}  & \includegraphics[width=.45\textwidth]{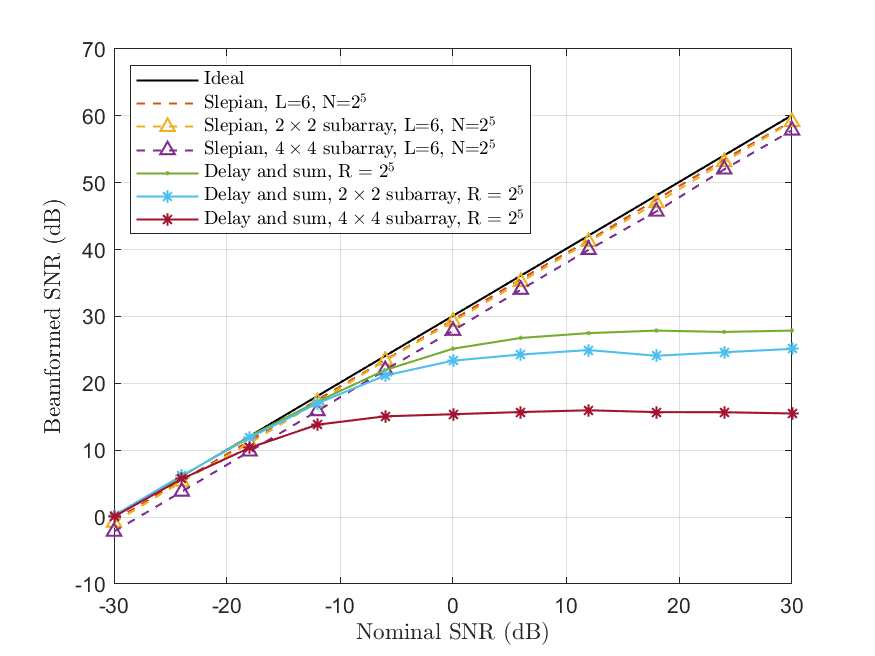}\\
		(a) 64 element ULA & (b) $32\times32$ element UPA
	\end{tabular}
	\caption{\small\sl A comparison of full scale delay and sum and Slepian beamforming to their subarray processing equivalents in a (a) 64 element ULA and a (b) $32\times32$ element UPA. Again we compare a series of nominal SNRs to the subsequent beamformed SNRs in order to observe the array gain with the ideal gain given by the black line.}
	\label{fig:sub_conv_bf_ag}
\end{figure}

Alongside both subarray configurations for both arrays the results shown in Figure~\ref{fig:sub_conv_bf_ag} also contain the full scale (e.g.\ no subarrays) results as a point of comparison. The ULA results in Figure~\ref{fig:sub_conv_bf_ag}(a) show that in the $2 \times 1$ subarray configuration there is almost no degradation in performance when using Slepian subspace beaforming and remains close to ideal. In contrast, delay and sum shows a faster performance roll off that is seen when operating at full scale. This roll-off becomes more pronounced in the $4 \times 1$ case, which is due to the signal nearly breaking the broadband threshold at the subarray level. Hence the narrowband beamforming stage incurs more distortion prior to filtering. However, there is also a sharp roll off in the Slepian performance when we switch to a $4 \times 1$ configuration. This is due to the fact that in this case $M'=16$, meaning $M'<d(\Omega T_1(\theta))=20$ and the spatial dimension is not sufficiently sampled.

Transitioning to the UPA case we see similar trends in the delay and sum results, which occur for the same reason. Again, the aperture of the larger subarray configuration is large enough to almost negate the ability to reasonably perform narrowband beamforming. The result is a drastic loss in performance when we transition form a $2\times 2$ to a $4 \times 4$ subarray configuration. On the other hand the Slepian subspace beamformer shows good results that track the ideal gain in both subarray configurations. This is because even for the larger subarray $M' = 64$ meaning the number of sub-beams far excedds the spatial subspace dimension of $d(\Omega T_1(\theta))=12$.

In conclusion the experiments show that subarray Slepian subspace beamforming can perform nearly as well as the full array case when the number of subarrays is sufficient. Performance of delay and sum is largely tied to the size of the subarray, with sharp performance degradation occurring when the subarray aperture is too large regardless of the number of subarrays. Furthermore, even when the subarray aperture is small there is a far wider performance gap between the subarray and full array versions of delay and sum than in the Slepian case.

\subsection{Bias, variance, and array gain}
\label{sec:biasvar}

As our broadband beamformer is a least-squares solver, we can systematically analyze its performance.  In this section, we characterize the performance of our least squares beamformer in \eqref{eq:lsconv} by showing how the bias and variance can be computed given the array geometry.  We show that the bias is very small while the variance yields the same (optimal) array gain.

For our analysis, we will compute the expected mean-squared error of the beamformed estimate when the incoming (demodulated) signal $s(t)$ is a Gaussian random process with zero mean and a flat power spectrum on $[-\Omega,\Omega]$.  The signal  is normalized so that its total expected energy on the observation interval $[0,T_N(\theta)]$ is $\Omega T_N(\theta)$:
\[
	\int_{0}^{T_N(\theta)}\E[|s(t)|^2]~\d{t} = \Omega T_N(\theta).
\]
The signal then has a Slepian decomposition with random coefficients
\begin{equation}
	\label{eq:flatslepian}
	s(t) = \sum_{k=1}^\infty \alpha_k\psi_k(t),
	\quad \alpha_k\sim\mathrm{Normal}(0,\lambda_k),
\end{equation}
where the $\alpha_k$ are independent and $\lambda_k$ is the $k$th largest eigenvalue of the integral operator in \eqref{eq:slepiankernelop}.  We observe $N$ array snapshots $\vybar$ as in \eqref{eq:yAtheta} in the presence of additive noise $\veta$, a random vector with zero mean and covariance matrix $\sigma^2\mId$ that is independent of $s(t)$.  We will use the least-squares technique in the previous section (with $\delta=0$) with $D_N(\theta)$ chosen for an appropriately small choice of $\epsilon$.

There are two sources of corruption in our forward model for the array samples in \eqref{eq:forwardmodel}.  The first is the standard additive signal-independent noise $\veta$.  The second is due to the \emph{model mismatch}.  Our least-squares recovery method only estimates the part of $s(t)$ spanned by the first $D_N(\theta)$ Slepian vectors; the remaining part of $s(t)$ still shows up in the samples and affects the least-squares estimate.  
When we sample element $m$ at time $t_n$, as in \eqref{eq:ymt}, we measure
\begin{align*}
	y_m(t) &= \e^{-j2\pi f_c\tau_m(\theta)}s(t_n-\tau_m(\theta)) + \eta_{m,n} \\
	&=  \e^{-j2\pi f_c\tau_m(\theta)}\sum_{k=1}^{D_N(\theta)}\alpha_k\psi_k(t_n-\tau_m(\theta)) + 
	\underbrace{\e^{-j2\pi f_c\tau_m(\theta)}\sum_{k>D_N(\theta)}\alpha_k\psi_k(t_n-\tau_m(\theta))}_{\text{mismatch}} + 
	\underbrace{\eta_{m,n}}_{\mathrm{noise}}.
\end{align*}
If we collect all the mismatch terms into a vector $\ve$, we have the observational model
\[
	\vybar = \mA(\theta)\valpha + \ve + \veta,
\]
where $\valpha = \{\alpha_1,\ldots,\alpha_{D_N(\theta)}\}$ in \eqref{eq:flatslepian}.

The estimated Slepian coefficients are $\widehat\valpha = \mA(\theta)^\pinv\vybar$.  Given these estimated Slepian coefficients  we take the estimated signal to be
\[
	\hat{s}(t) = \sum_{k=1}^{D_N(\theta)}\widehat\alpha_k\psi_k(t).
\]
We can break down the mean-squared error into two bias terms and one variance term:
\begin{align}
	\nonumber
	\int_0^{T_N(\theta)} \E|s(t)-\hat{s}(t)|^2~\d{t} &= 
	\sum_{k>D_N(\theta)}\E|\alpha_k|^2 +
	\sum_{k=1}^{D_N(\theta)} \E|\alpha_k-\widehat\alpha_k|^2 \\
	\nonumber
	&= \sum_{k>D_N(\theta)}\E|\alpha_k|^2 + \|\mA(\theta)^\pinv\ve\|_2^2 + \|\mA(\theta)^\pinv\veta\|_2^2
	\quad\text{($s(t)$ and $\veta$ are independent)} \\
	\label{eq:threeerrors}
	&= \underbrace{\sum_{k>D_N(\theta)}\E|\alpha_k|^2}_{\text{truncation bias}} + 
	\underbrace{\trace(\mA(\theta)^\pinv\E[\ve\ve^\H]\mA(\theta)^{\pinv\H})}_{\text{mismatch bias}} + 
	\underbrace{\sigma^2\trace((\mA(\theta)^\H\mA(\theta))^{-1})}_{\text{variance}}
\end{align}
A bound on the truncation bias follows directly from our definition of $D_N(\theta)$:
\begin{equation}
	\label{eq:truncationbias}
	\sum_{k>D_N(\theta)}\E|\alpha_k|^2 \leq \epsilon\Omega T_N(\theta).
\end{equation}

The mismatch bias and variance terms can be computed from the array geometry.  To do this, we can rewrite the mismatch bias term as follows.  Let  $\va_k(\theta) = \{\e^{-j2\pi f_c\tau_m(\theta)}\psi_k(t_n-\tau_m(\theta))\}_{m,n}$ be the vector that contains the (modulated) samples of $\psi_k$ at all sample times and delay offsets (note that $\va_k(\theta)$ for $k=1,\ldots,D_N(\theta)$ make up the rows of $\mA(\theta)$).  The entries of $\va_k(\theta)$ are naturally indexed by $\ell=1,2,\ldots,MN$ where each value of $\ell$ can be put in one-to-one correspondence to a pair $(m,n)$.  As such, we denote $\tau_\ell = t_n-\tau_m(\theta)$.  Note that
\[
	\sum_{k=1}^\infty\lambda_k\va_k(\theta)\va_k(\theta)^\H = \mB(\theta)
	,\quad\text{where}\quad
	B(\theta)[\ell,\ell'] = \frac{\sin(\Omega(\tau_{\ell}-\tau_{\ell'}))}{\pi(\tau_\ell-\tau_{\ell'})},
\]
and so
\begin{align*}
	\trace(\mA(\theta)^\pinv\E[\ve\ve^\H]\mA(\theta)^{\pinv\H}) &= \trace\left(\mA(\theta)^\pinv\left(\sum_{k>D_N(\theta)}\E[|\alpha_k|^2]\va_k(\theta)\va_k(\theta)^\H\right)\mA(\theta)^{\pinv\H}\right) \\
	&= \trace\left(\mA(\theta)^\pinv\left(\mB(\theta) - \sum_{k=1}^{D_N(\theta)}\lambda_k\va_k(\theta)\va_k(\theta)^\H\right)\mA(\theta)^{\pinv\H}\right)
\end{align*}
This quantity, along with the variance $\sigma^2\trace((\mA(\theta)^\H\mA(\theta))^{-1})$, can be computed by evaluating a finite number of Slepian basis functions at a finite number of places determined by the array geometry and angle of arrival.  This can be done to high precision using any number of methods that exist in the literature (see for example \cite{osipov14on,bremer22on}).  We have included tabulations of these quantities in Table~\ref{tab:bias_var_ula} for the experiments considered in Section~\ref{sec:conventionalexperiments}.
\begin{table}[h]
    \centering
    \begin{tabular}{|c|c|c|c|c|c|}
        \hline
        \multicolumn{6}{|c|}{Uniform linear array bias and variance quantities}\\
        \hline
         Error type & $L=0$ & $L=2$ & $L=4$ & $L = 6$ & $L=8$  \\
         \hline
          Truncation bias & 0.0021 & $1.99\times 10^{-4}$ & $8.04 \times 10^{-6}$ & $2.09 \times 10^{-7}$ & $3.96 \times 10^{-9}$\\
          \hline
          Mismatch bias & 0.0012 & $2.16 \times 10^{-6}$ & $2.13 \times 10^{-8}$ & $2.92 \times 10^{-10}$ & $4.27 \times 10^{-12}$\\
          \hline
          {\small$\trace((\mA(\theta)^\H\mA(\theta))^{-1})$} & 0.037 & 0.042 & 0.046 & 0.05 & 0.053\\
          \hline
    \end{tabular}
    \caption{\small \sl Computation of the truncation bias, mismatch bias, and variance multiplier for the ULA scenario described in Section~\ref{sec:conventionalexperiments}. The truncation and mismatch bias have been normalized by $\Omega T_N(\theta)$.}
    \label{tab:bias_var_ula}
\end{table}

\begin{figure}[h]
	\centering
	\begin{tabular}{cc}
		\includegraphics[width=.45\textwidth]{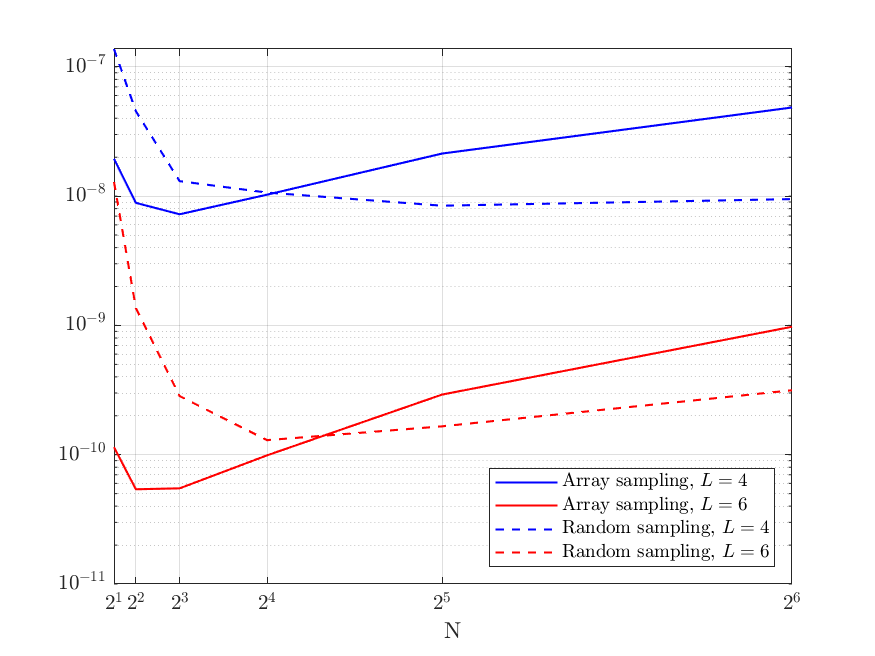}  & \includegraphics[width=.45\textwidth]{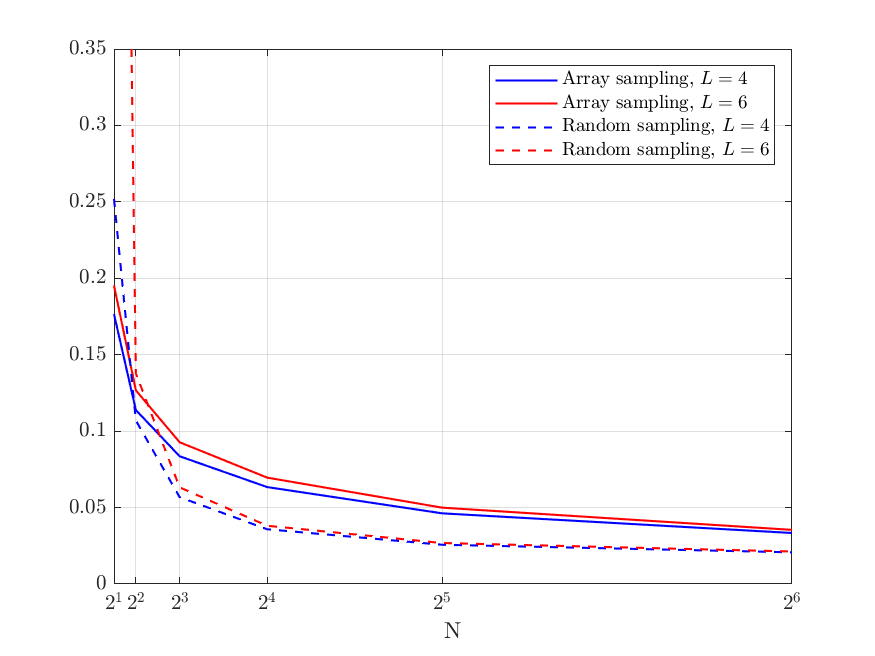}\\
		(a) Mismatch bias & (b) Variance
	\end{tabular}
	\caption{\small\sl A comparison of the (a) mismatch bias and (b) variance for sampling patterns generated at random and from the ULA described in Section~\ref{sec:conventionalexperiments}. We compare two choices of $L$ and vary the number of snapshots we observe to get a better idea of how they behave under each samplng scheme.}
	\label{fig:bias_var}
\end{figure}

We can also get a rough idea of what to expect from these terms using recent results from random matrix theory.  Instead of considering the sampling pattern induced by the $\{t_n-\tau_m(\theta)\}$, we can ask what the variance and mismatch error might look like for a  ``generic'' set of $MN$ samples on the interval $[0,T_N(\theta)]$, where $\tau_\ell, ~\ell=1,\ldots,MN$ are drawn uniformly at random.  Of course the sample locations are not drawn at random in practice, but as the calculations below and the results in Figure~\ref{fig:bias_var} show, the behavior is qualitatively very similar.  For a fixed angle of arrival $\theta$, we will use the same model for $s(t)$ as in \eqref{eq:flatslepian} with $D_N(\theta)$ and $T_N(\theta)$ chosen accordingly.

With the $\tau_\ell$ drawn randomly, we form, in a similar manner as before,\footnote{We are omitting the $\theta$ for the $\va_k$ because the sample locations are random and not determined by the angle of arrival.  We are also omitting the phase term $\e^{-j2\pi f_c\tau_m(\theta)}$ as it does not impact the discussion at all at this point.} the vectors $\va_k\in\C^{MN}$ with $\va_k[\ell] = \psi_k(\tau_\ell)$ and construct $\mA$ using the $\{\va_k\}_{k=1}^{D_N(\theta)}$ as rows.

With this model, we can use standard tools to bound both the variance and the mismatch.  We forgo a detailed argument here, both for the sake of space and because there are multiple examples in the literature of very similar results (including \cite{bass04ra}, \cite{rauhut12sp}, \cite{cohen13st}, and \cite[Prop.\ 5.2]{romberg22st}).  The key point is that $\mA$ is a random matrix with independent rows and bounded entries; applying \cite[Thm.\ 5.41]{vershynin12in} we know that for $MN\gtrsim D_N(\theta)\log(D_N(\theta))$,
\[
(1-\delta)MN\leq\sigma^2_{\mathrm{min}}(\mA) \leq \sigma^2_{\mathrm{max}}(\mA)\leq (1+\delta)MN,
\]
with high probability, where $\delta$ is a small constant ($\sigma_{\text{min}}^2(\mA)$ is the smallest squared singular value of $\mA$ and smallest eigenvalue of $\mA^\H\mA$).  This means that the variance term in \eqref{eq:threeerrors} behaves like
\begin{equation}
	\label{eq:randomarraygain}
	\sigma^2\trace\left((\mA^\H\mA)^{-1}\right) \lesssim \sigma^2\frac{D_N(\theta)}{MN}.
\end{equation}

We can also rewrite the mismatch bias as
\begin{align*}
 	\trace(\mA(\theta)^\pinv\E[\ve\ve^\H]\mA(\theta)^{\pinv\H})  
 	&=\trace\left(\mA(\theta)^\pinv\left(\sum_{k>D_N(\theta)}\E[|\alpha_k|^2]\va_k(\theta)\va_k(\theta)^\H\right)\mA(\theta)^{\pinv\H}\right) \\
 	&= \sum_{k>D_N(\theta)}\lambda_k \|\mA(\theta)^\pinv\va_k(\theta)\|_2^2 \\
   	&\leq \epsilon\Omega T_N(\vtheta)\, MN\mu/\sigma^2_{\mathrm{min}}(\mA) \\
	&\lesssim \epsilon\Omega T_N(\theta),
 \end{align*}
which is on the same order as the truncation bias in \eqref{eq:truncationbias} and ultimately negligible as we can choose $\epsilon$ very small.  We have used $\mu = \sup_{k}\|\psi_k(t)\|_\infty$ above.

\textbf{Array gain.}  The dependence of the variance on $M$ and $N$ tells us what the ``array gain'' is for our beamformer.  For a standard conventional beamformer, implemented using delay-and-sum, the variance  scales like $\sigma^2/M$: increasing the number of array elements not only increases the spatial selectivity but also cancels noise in the signal estimate.  Our variance estimate in \eqref{eq:randomarraygain} behaves similarly.  Since $D_N(\theta) = 2\Omega T_1(\theta) + L - 1 + N$, we have that $\sigma^2D_N(\theta)/MN\rightarrow \sigma^2/M$ as $N$ gets large.  Thus we asymptotically recover the full array gain.

Note that for delay-and-sum to achieve the array gain of $1/M$ it needs to use sinc intepolation filters with infinite extent.  In practice, these fractional delays have to be implemented with finite-length filters, and this introduces bias into the estimate of the signal.  The experiments in the last section suggest that this bias is significantly worse than the truncation and mismatch biases discussed for our technique above. 

\subsection{Computational complexity}
\label{sec:compcomplex}

The number of computations needed to implement the least-squares approach to beamforming outlined above is not too different than what is needed for standard delay and sum.  For an $M$-element array processing a batch of $N$ snapshots, the pseudo-inverse in \eqref{eq:pseudoinv} is $D_N(\theta)\times MN$.  As for moderate $N$ we will have $D_N(\theta)\sim N$, the computational complexity is $O(MN)$ per sample. For a fixed angle the pseudo-inverse does not change batch-to-batch. Hence it only needs to be calculated once and can be precomputed. 

This can be improved by taking advantage of the fact that the snapshots $\vy[n]$ are themselves low-dimensional \cite{delude2022bro}.  As discussed in Section~\ref{sec:slepian}, each $\vy[n]$ (approximately) lies in a subspace of dimension $D_1(\theta)$, meaning we can write $\vy[n] \approx \mU\vbeta_n$.  This subspace is also (approximately) in the column space of $\mA_n(\theta)$; we can write $\mA_n(\theta) \approx \mU\mC_n$, where $\mC_n$ is $D_1(\theta)\times D_N(\theta)$.  Thus we can re-cast the least-square problem in \eqref{eq:lsconv} as
\[
	\minimize_{\valpha\in\C^{D_N(\theta)}}~\sum_{n=1}^N\|\vbeta_n-\mC_n\valpha\|_2^2 + \delta\|\valpha\|_2,
\]
which can  be solved by applying a $D_N(\theta)\times D_1(\theta)N$ matrix to the concatenation $\{\vbeta_n\}$.  The per-sample computational complexity in this case is $O(D_1(\theta)M + D_1(\theta)N)$, where the first term is the cost of ``encoding'' the snapshots (computing $\beta_n=\mU^\H\vy[n]$) and the second term is the cost of applying the pseudo-inverse that solves the least squares problem.  Since in general, $D_1(\theta)\ll M$ and we will want to choose $N$ as large as possible to get the best array gain, the computational savings using snapshot encoding can be significant.

\begin{figure}[h]
    \centering
    \includegraphics[width = .5\textwidth]{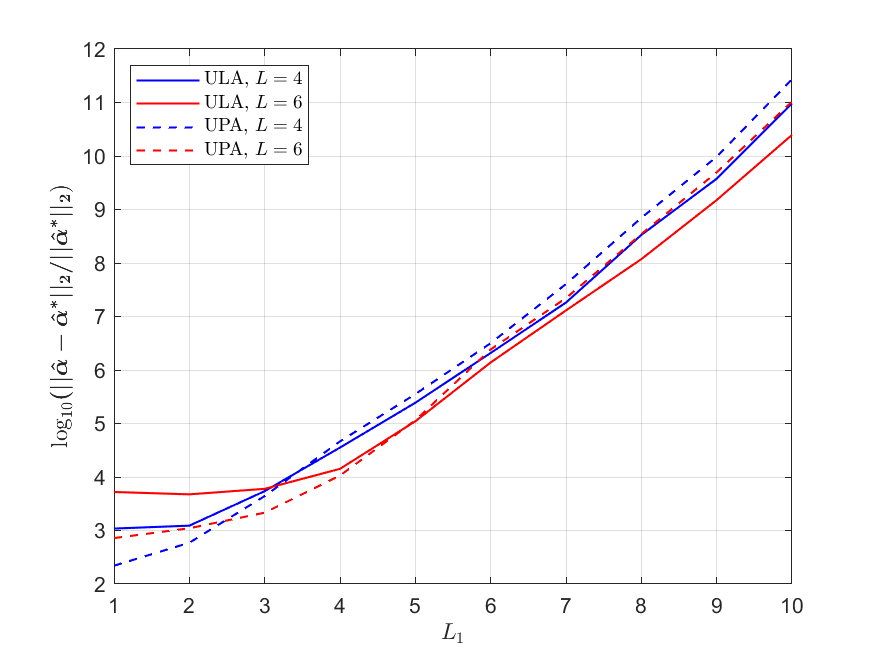}
    \caption{\small \sl The plot above shows the difference (on a $\log_{10}$ scale) between the full non-encoded least squares solution $\widehat{\valpha}^*$ and the encoded ``efficient" least squares solution $\widehat{\valpha}$ for the test scenarios examined in Section~\ref{sec:conventionalexperiments}. $L_1$ is the per snapshot equivalent of the parameter $L$, e.g.\ the dimension of the subspace is chosen to be $\lceil 2 \Omega T_1(\theta) \rceil + L_1$. Naturally as we increase $L_1$ the approximation of improved. For very modest choices in $L_1$ we achieve a solution to within several digits of accuracy to the full least squares solution.}
    \label{fig:efficient_ls_alpha_diff}
\end{figure}

The snapshot encoding discussed in the previous paragraph also introduces another source of error, as hinted at by the two parenthetical instances of ``approximately''.  Replacing $\vy[n]$ with $\mU\beta_n$ and $\mA(\theta)$ with $\mU\mC_n$ introduces truncation error that can be analyzed in a very similar way to what was done in Section~\ref{sec:biasvar}.  We do expect, however, this additional bias to be very small.  The results in Figure~\ref{fig:efficient_ls_alpha_diff} show that almost nothing is lost empirically.

Much of the time, $D_1(\theta)$ can be treated as a small constant; the broadband experiments in Section~\ref{sec:conventionalexperiments} have $D_1(\theta)=15$ for the UPA case and realize essentially the full array gain for $N=32\ll M$.  The computational complexity is thus similar to the $O(RM)$ per sample for a delay-and-sum implementation with $R$-tap delay filters while the estimation performance is significantly better.

At first glance, it is somewhat alarming that the per-sample computational complexity scales with the batch size $N$.  We will see in Section~\ref{sec:streaming}, however, that this least-squares estimation approach can be solved for arbitrarily large time intervals by using a streaming algorithm that breaks the signals into batches of size $N\sim D_1(\theta)$ and uses a small number of matrix operations per batch.

\section{Adaptive beamforming}
\label{sec:adaptive}

It is often the case that the signal we are beamforming is not the only signal impinging on the array. Other signals, which we term ``interferers," are assumed to be present in most applications. These interferers can arise from several sources such as natural phenomina (e.g.\ clutter) or directly from adjacent transmitters. Regardless of the source, adaptive beamforming uses statistics on the received signals to simultaneously beamform the signal while nulling out interferers. How we obtain these statistics, namely how we estimate the covariance, is itself a topic of research. Therefore as we move forward describing these adaptive methods we will focus on the case where the covariance is known or has been sufficiently estimated.

Our least-squares approach naturally extends to adaptive beamforming, and in this section we will discuss two methodologies.  In the first, we show how to null an interferer coming from a known direction.  In the second, we show that the estimation process can be adapted to a known array covariance matrix, giving us a natural extension to classic MVDR and LCMV beamformers.

Operationally, these methods amount to replacing the pseudo-inverse in \eqref{eq:pseudoinv} with a different $D_N(\theta)\times MN$ matrix.  We require no pre-steering or other auxiliary filtering.  This allows the performance of the proposed methods to be virtually independent of the magnitude of the interferers (given that the signals have been digitized faithfully).

\subsection{Nulling an interferer}
\label{sec:slepiannulling}

\begin{figure}[h]
    \centering
    \includegraphics[width = .5\textwidth]{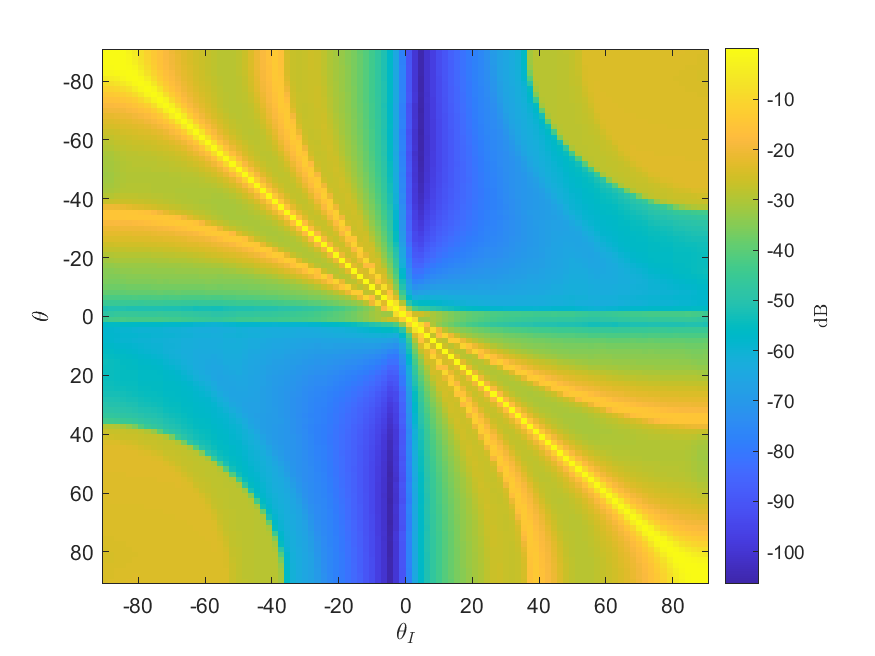}
    \caption{\small \sl A plot of the additional bias incurred by placing a null for the ULA described in Section~\ref{sec:conventionalexperiments}. The signal and interferer subspace truncation threshold is set using $D_N(\theta)$ and $D_N(\theta_I)$ respectively. We observe that in general the additional bias is small, but as the interferer comes closer to the signal of interest the bias increases.}
    \label{fig:slepian_nulling_ula_bias}
\end{figure}

To null out an interferer at angle $\theta_I$ while estimating a signal incident from angle $\theta$, we remove the effects of the interferer from the array samples $\vybar$ using a linear projection
\[
	\tilde{\vybar} = \mP^\perp(\theta_I)\vybar, \quad\text{where}~~
	\mP^\perp(\theta_I) = \mId - \mP(\theta_I) = \mId - \mA(\theta_I)\left(\mA(\theta_I)^\H\mA(\theta_I)\right)^{-1}\mA(\theta_I)^\H,
\]
and then apply the estimation procedure from \eqref{eq:pseudoinv}
\[
	\widehat\valpha = \mA(\theta)^\pinv\tilde{\vybar} = \mA(\theta)^\pinv\mP^\perp(\theta_I)\vybar.
\]
This has the effect of completely (to within the truncation parameter $\epsilon$ used to choose $D_N(\theta_I)$) removing anything coming from angle $\theta_I$; if $\vybar = \mA(\theta_I)\valpha_I$  for any $\valpha_I\in\C^{D_N(\theta)}$, then $\mP^\perp(\theta_I)\vybar = \vzero$.  The nulling also introduces additional bias into the estimate, as we are losing any part of the signal that ``bleeds'' into the subspace spanned by the columns of $\mA(\theta_I)$.  If $\vybar = \mA(\theta)\valpha + \mA(\theta_I)\valpha_I$, then
\[
	\valpha-\widehat\valpha = \mA(\theta)^\pinv\mP(\theta_I)\mA(\theta)\valpha.
\]
Thus this additional bias is a function of the spectral properties of $\mA(\theta)^\pinv\mP(\theta_I)\mA(\theta)$.  For example, if $\valpha$ follows the model in \eqref{eq:flatslepian}, then we have $\E[\|\valpha-\widehat\valpha\|_2^2] = \trace(\mZ\mLambda\mZ^\H)$, where $\mLambda$ is a diagonal matrix of $\lambda_1,\ldots,\lambda_{D_N(\theta)}$ and $\mZ = \mA(\theta)^\pinv\mP(\theta_I)\mA(\theta)$.  While this expression is complicated, it is again completely determined by the array geometry, $\Omega$, and $f_c$. In our experiments below (See Figure~\ref{fig:slepian_nulling_ula_bias}), we compute this quantity explicitly and show how it varies with $\theta$ and $\theta_I$.

\subsection{Adapting to known covariance: MVDR and LMCV}
\label{sec:slepianmvdr}

As previously alluded to, our least-squares approach also allows us to naturally extend covariance-based adaptive beamforming techniques to broadband.  We will discuss below the extensions to the MVDR and the LCMV beamformers, but the essential idea is to design for a matrix of weights instead of a single vector.

In narrowband adaptive beamforming, the estimate of the signal is formed sample-by-sample as $\hat{s}[n] = \vw^\H\vy[n]$.  For the MVDR beamformer, the weights $\vw$ are chosen to minimize the output power while ensuring a ``distortionless response'' in a given direction $\theta$ through the constraint $\vw^\H\va(\theta) = 1$ ($\va(\theta)$ is the steering vector used in the conventional beamformer).  With $\mR = \E[\vy[n]\vy[n]^\H]$ as the $M\times M$ covariance matrix of the array snapshots, the weights are the solution to 
\[
	\minimize_{\vw\in\C^N}~\vw^\H\mR\vw\quad\text{subject to}~~\vw^\H\va(\theta) = 1,
\]
which has closed-form solution $\hat\vw = (\va(\theta)^\H\mR^{-1}\va(\theta))^{-1}\mR^{-1}\va(\theta)$.

In the broadband case, we are mapping one or more array snapshots to multiple signal parameters (Slepian coefficients).  This mapping is given by a $D_N(\theta)\times MN$ matrix (e.g.\ the psuedo-inverse $\mA(\theta)^\pinv$ used for the conventional beamformer above).  To extend the MVDR to broadband, we form the $MN\times MN$ covariance matrix $\mRbar = \E[\vybar\vybar^\H]$ and solve
\begin{equation}
	\label{eq:bbmvdr}
	\minimize_{\mW}~\trace\left(\mW\mRbar\mW^\H\right)\quad\text{subject to}~~\mW\mA(\theta) = \mId.
\end{equation}
This is a linearly-constrained convex quadratic program, and (like all such programs) has a closed form solution
\[
	\widehat\mW = \left(\mA(\theta)^\H\mRbar^{-1}\mA(\theta)\right)^{-1}\mA(\theta)^\H\mRbar^{-1}.
\]
This program has the same interpretation as in the narrowband case; we are minimizing the energy of the estimated signal (given by $\|\valpha\|_2^2$, since the $\{\psi_k\}$ are an orthobasis) while preserving everything coming from direction $\theta$ (meaning $\mW$ is a left inverse of $\mA(\theta)$).  Given this matrix of weights, we form the estimate of the expansion coefficients using $\widehat\valpha = \mW\vybar$.

We can add  constraints to \eqref{eq:bbmvdr} to get the LMCV beamformer, just as in the narrowband case.  For example, if we want to focus the beam in direction $\theta$ while explicitly nulling an interferer in direction $\theta_I$, we simply add the linear constraint $\mW\mA(\theta_I) = 0$.  This is still a linear constrained quadratic program and so solving for the corresponding weights $\mW$ remains tractable.

\subsection{Numerical experiments: Comparison to adaptive beamforming}
\label{sec:adaptiveexperiments}


To see how the filter based adaptive approaches compare against our Slepian ones we examine the same array described in Section~\ref{sec:conventionalexperiments} but add a randomly generated interfering signal to the test scenario. This signal resides at the same center frequency and possesses the same bandwidth but is incident at a reasonably separated angle from the signal of interest.\footnote{Here, by ``reasonable'' we mean that if we extended the ULA or UPA infinitely and calculated the 1D or 2D discrete-time Fourier transform across a snapshot, then the respective signals' occupied bands do not overlap.} The interferer arrives at $\theta = \pi/6$ for the ULA and $\theta = (-\pi/4,\pi/2)$ for the UPA. The signal, interferer, and noise covariance matrix are assumed to be known a priori. Hence we examine the ideal case where statistics on the signal are known. In our experiments we vary both the SNR and the signal to interferer ratio (SIR), and again average over 50 trials. 

The Slepian MVDR/MPDR\footnote{Minimum power distortionless response (MPDR) and MVDR have slightly non-standard definitions in \cite[Ch.\ 6]{vantrees02op} with the distinction being whether the signal and noise/interferer covariance matrix can be decoupled.} and Slepian nulling beamformers are designed as described in Section~\ref{sec:slepianmvdr} and Section~\ref{sec:slepiannulling} respectively. The traditional filter based MVDR/MPDR beamformers are designed as described in \cite[Ch.\ 6]{vantrees02op}, and we examine the ideal case of perfect pre-steering (e.g.,\ infinite length fractional delay filters) and imperfect finite length filter pre-steering using 64-tap sinc interpolators. The LCMV beamformer was designed using eigenvector constraints as described in \cite[Ch.\ 2]{liu2010wi}, and implicitly requires no pre-steering. Each filter is length $R=2^5$ and the Slepian MVDR/MPDR beamformer operates on a signal block containing $N=2^5$ snapshots. 

We note that each of these methods with the exception of Slepian nulling requires a covariance matrix inversion. This matrix is extremely large in the $32 \times 32$ array case making direct inversion difficult. However, since we have access to the signal and interferer covariance matrix, and they are known to be numerically low rank, we can efficiently form a good low rank approximation to both. Inverting these low rank approximations remains computationally tractable, and we utilize this technique in our experiments. Further details on this methodology can be found in Appendix~\ref{apx:covaprx}.

The results shown in Figure~\ref{fig:adaptive_bf_ag_vs_snr} fix the SIR of the interferer to $-30$ dB and varies the nominal SNR, again we compare the nominal SNR to the beamformed SNR. We observe that the performance of filter-based MVDR/MPDR with ideal pre-steering tracks the ideal array gain nearly perfectly. However, when imperfect pre-steering is used we see a performance roll-off similar to those observed in the conventional beamforming experiments. LCMV, which does not require presteering, has a relatively constant array gain loss in some SNR regimes and outperforms MVDR/MPDR in some regimes. However, the performance degrades as the SNR increases likely due to self-cancelation as the eigenvector constraints become insufficient at representing the signal across all possible frequencies. In contrast to the filter based methods the Slepian MPDR and nulling methods (which implicitly require no pre-steering) track the ideal array gain across all tested SNRs. Only the Slepian nulling shows some degradation in performance at the highest of SNRs, but not nearly as much as filter based MVDR/MPDR.
\begin{figure}[h]
	\centering
	\begin{tabular}{cc}
		\includegraphics[width=.45\textwidth]{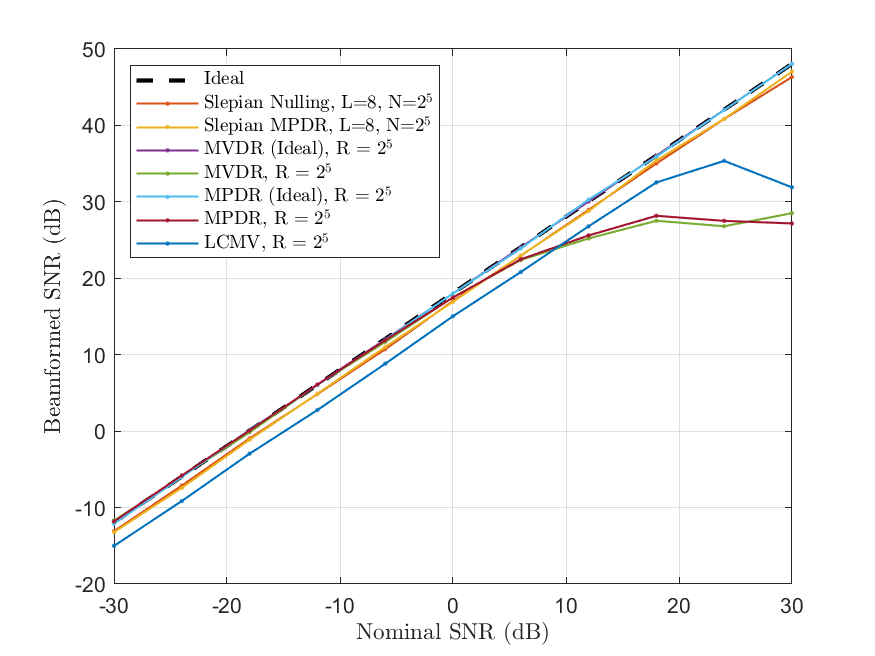}  & \includegraphics[width=.45\textwidth]{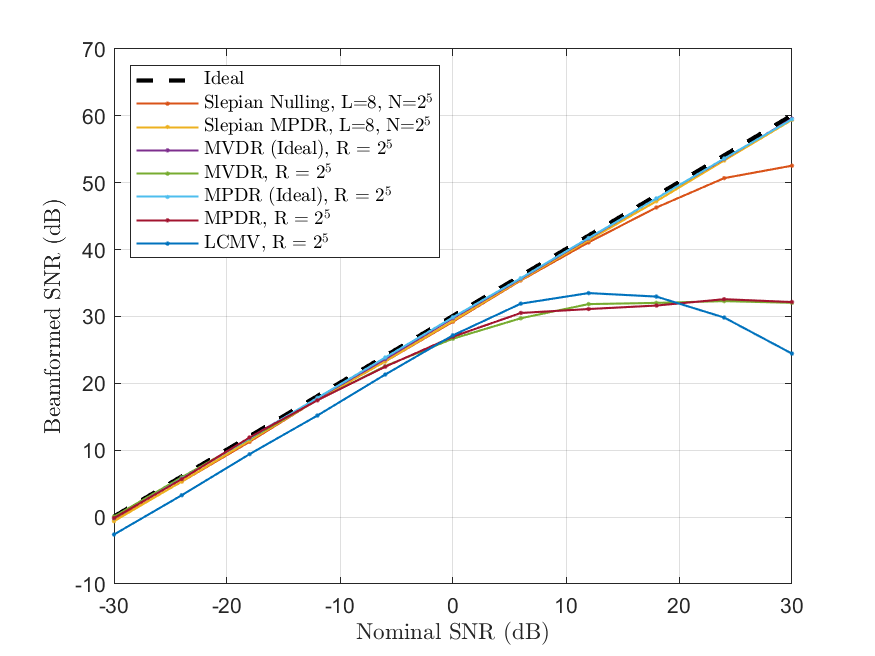}\\
		(a) 64 element ULA & (b) $32\times32$ element UPA
	\end{tabular}
	\caption{\small\sl A comparison of classical adaptive methods with pre-steering (MVDR and MPDR), classical methods without pre-steering (LCMV), Slepian subspace equivalent of MPDR, and pseudo-adaptive Slepian methods (e.g.,\ Slepian beamforming with a null placed) for a (a) 64 element ULA and a (b) $32\times32$ element UPA. An interfering signal with an SIR of $-30$ dB is incident at a reasonably separated angle in both cases.}
	\label{fig:adaptive_bf_ag_vs_snr}
\end{figure}

\begin{figure}[h]
	\centering
	\begin{tabular}{cc}
		\includegraphics[width=.45\textwidth]{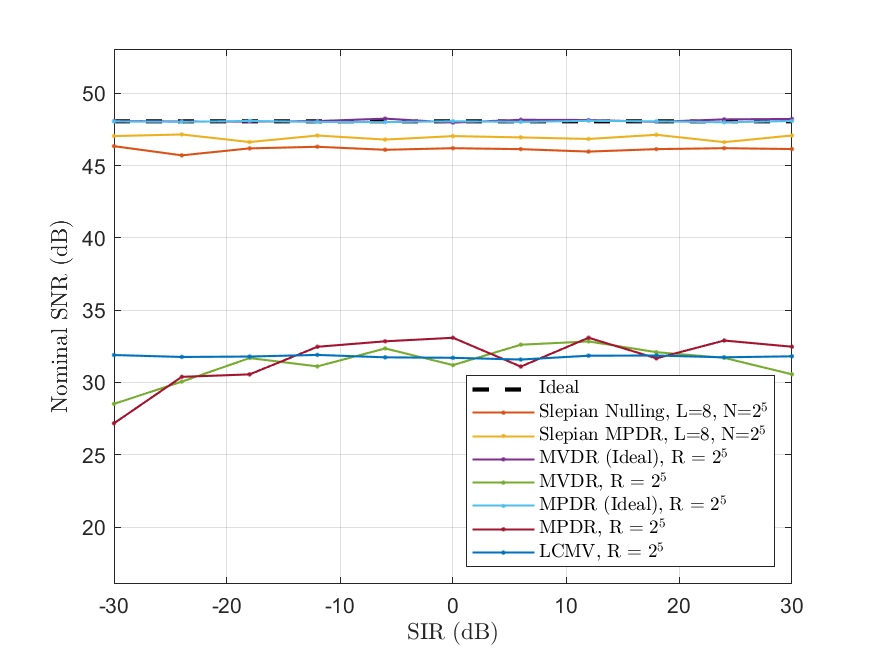}  & \includegraphics[width=.45\textwidth]{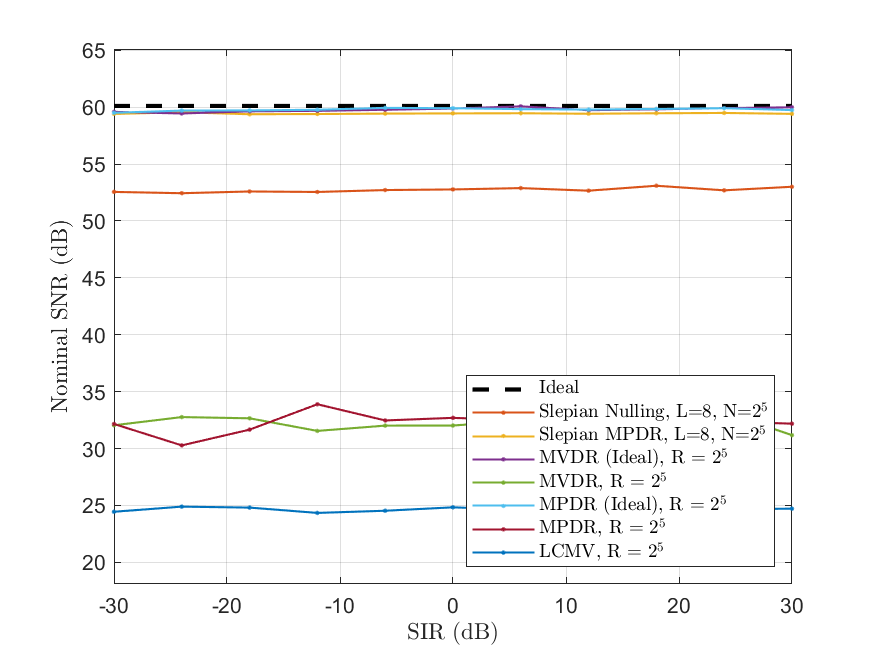}\\
		(a) 64 element ULA & (b) $32\times32$ element UPA
	\end{tabular}
	\caption{\small\sl A comparison of classical adaptive methods with pre-steering (MVDR and MPDR), classical methods without pre-steering (LCMV), Slepian subspace equivalent of MPDR, and pseudo-adaptive Slepian methods (e.g.\ Slepian beamforming with a null placed) for a (a) 64 element ULA and a (b) $32\times32$ element UPA. A series of SIRs for an interfering signal incident at a reasonably separated angle are compared to the subsequent array gains of each method. The SNR is fixed to $30$ dB for all test cases.}
	\label{fig:adaptive_bf_ag_vs_sir}
\end{figure}

As a complementary comparison, Figure~\ref{fig:adaptive_bf_ag_vs_sir} fixes the SNR to $30$ dB and varies the SIR. All methods seem to show fairly consistent interferer cancellation across the tested SIR range. However we do see a roll-off for MVDR/MPDR as the SIR exceed $-20$ dB for the ULA. This trend is not seen in the UPA case, which is likely due to it having a substantially larger number of elements allowing for greater interferer cancellation.

From these experiments we conclude that the Slepian MVDR and nulling techniques devised in the previous sections are able to effectively cancel out strong interferers and preserve array gain. When compared to an idealized filter based MVDR/MPDR we see that the Slepian techniques slightly under perform in comparison. However, when we take into account realistic non-idealized pre-steering the Slepian techniques dominate MVDR, MPDR, and LCMV beamforming.

\section{Streaming reconstruction}
\label{sec:streaming}

In Section~\ref{sec:conventional} and \ref{sec:adaptive} above, we showed how the Slepian coefficients, which in turn can be translated into equispaced Nyquist samples, of a beam in a particular direction $\theta$ over a time interval of length $T_N(\theta)$ can be estimated from $N$ array snapshots.  The estimation procedure consisted of solving a least-squares problem by applying a fixed matrix that mapped the $MN$ array samples to $D_N(\theta)$ Slepian coefficients then generating the desired samples from these coefficients.  We also saw that as the number of snapshots $N$ increased, the array gain improved.  But the cost of applying the matrix also increases, scaling as $O(N^2)$ as does the latency --- the signal cannot be estimated until all $N$ snapshots have been collected.

\begin{figure}
	\centering
	\includegraphics[height=1.75in]{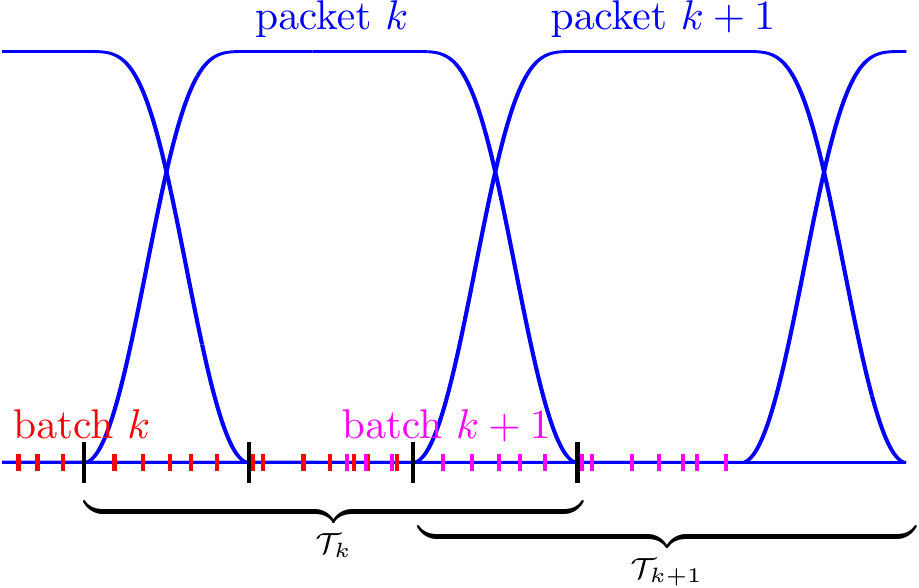}
	\caption{\small\sl Sketch of the key quantities in the streaming reconstruction formulation.  The signal is divided into overlapping packets, each equipped with its own basis expansion.  The array snapshots are divided into batches that overlap at most two of the signal packet supports.  After observing sample batch $K+1$, Algorithm~\ref{alg:streaming} shows how to form an initial estimate of the signal in packet $K$ and then update the estimates for all previous packets.}
	\label{fig:samplebatches}
\end{figure}

In this section, we show how the least-squares problem in \eqref{eq:lsconv} can be solved in an online manner by adapting recent work in \cite{romberg22st}.  The signal is divided into  overlapping \emph{packets}; while the array snapshots are divided into similarly sized \emph{batches}.  The sizes of the packets and the batches are adjustable with the only constraint being that the temporal extent of each batch can overlap with the temporal extent of at most two packets (see Figure~\ref{fig:samplebatches}).  When a snapshot batch arrives, we estimate the corresponding beam packet and then backtrack, updating the estimates of beam packets in the past.  The result is a solver that always has a current optimal estimate of the beam signal in hand and can (approximately, but with provable accuracy) solve the least-squares problem \eqref{eq:lsconv} for an arbitrarily large number of array snapshots.

We now make this formulation precise.  To streamline the exposition below, we will drop the dependence of the parameters on the angle of arrival $\theta$.  We divide the array samples into batches consisting of $N$ snapshots, so each batch contains $MN$ samples of the signal.  We will index the batches by $k$, and denote the samples in batch $k$ as $\vybar_k$ and their corresponding sample times as $\vtaubar_k$.  We will assume that each batch is the same size and has extent in time. 

The signal $s(t)$ to estimate is divided into packets that are supported locally in time.  These packets are also indexed by $k$.  Each packet is supported on a time interval $\setT_k$; consecutive intervals overlap with each other (so $\setT_k\cap\setT_{k-1}$ is not empty) but not with other intervals (so $\setT_k\cap\setT_{k-2}$ is empty).  We also ensure (through the choice of $N$ and the $\setT_k$) that that each batch of samples ``touches'' time intervals $\setT_k$ and $\setT_{k-1}$  and only these time intervals: $\vtaubar_k\subset(\setT_k\cup\set T_{k-1})\backslash(\setT_{k-2}\cup\setT_{k+1})$.  These constraints are shown in Figure~\ref{fig:samplebatches}.

Inside of packet $k$, we represent the signal using basis functions $\{\psi_{k,d}\}_{d=1}^D$.  The $\psi_{k,d}$ are Slepian functions that have been projected into overlapping orthogonal blocks, forming a lapped orthogonal transform \cite{malvar92si,mallat09wa}.  They are essentially Slepian functions that taper off smoothly towards zero at the edge of the interval $\setT_k$; their envelope is what is depicted in Figure~\ref{fig:samplebatches}.  They are carefully constructed so that $\{\psi_{k,d}(t),~k\in\Z,d\geq 1\}$ is an orthonormal sequence in $L_2(\R)$.  They also enjoy the same spectral concentration properties as the standard Slepian functions in that choosing $D\approx 2\Omega T$, where $T$ is the length of the interval $\setT_k\backslash\setT_{k-1}$ (details about this construction can be found in \cite{romberg22st}).  Signal packet $k$ is specified though its expansion coefficients $\valpha_k\in\C^D$.

We can now formulate the streaming analog to the least-squares beamformer in \eqref{eq:lsconv}.  We define the two $MN\times D$ matrices $\mA,\mB$ as
\begin{align*}
	A_{\ell,d} &= \psi_{k,d}(\tau_\ell), \\
	B_{\ell,d} &= \psi_{k-1,d}(\tau_\ell),~~~\ell=1,\ldots,MN,~~d=1,\ldots,D,
\end{align*}
where the $\tau_\ell$ are the entries in $\vtaubar_k$.  Note that the way we have defined the sample batches and the packet intervals means that  $\mA$ and $\mB$ do not depend on $k$.  Below, we will also use the matrix $\mE = \mA^\H\mB$.

At every index $K$, we want to solve the optimization program\footnote{We can simply fix $\valpha_{-1}=\vzero$ to make every term in this sum fully defined.}
\begin{equation}
	\label{eq:lsK}
	\minimize_{\{\valpha_0,\ldots,\valpha_K\}}~\sum_{k=0}^{K} \left\|\mA\valpha_k + \mB\valpha_{k-1} - \vybar_k\right\|_2^2 + \delta\sum_{k=0}^K\|\valpha_k\|_2^2.
\end{equation}
Because the terms inside the sum above couple consecutive packets together, the solution to this program changes with every increment to $K$; that is, the optimal estimate $\valpha_k$ for fixed $k$ will update every time a new batch arrives and $K$ increases.

Algorithm~\ref{alg:streaming} solves \eqref{eq:lsK} in a streaming manner.  After samples in batch $k$ are observed, the algorithm forms an initial estimate of the signal in packet $k$, then backtracks updating the signal estimate in all packets prior to $k$.  In practice, these estimates settle after a small number of updates, so limiting the backtracking step to a buffer of some fixed length $B$ will produce an extremely accurate approximation of the true solution (analysis of this can be found in \cite{romberg22st}). Our numerical below use a buffer length of $B=5$.
\begin{algorithm}
	\begin{algorithmic}
		\State Observe array sample batches $\vybar_0,\vybar_1$
		\State $\mQ_0 \gets  \mA^\H\mA + \mB^\H\mB + \delta\mId$
		\State $\mU_0\gets \mQ_0^{-1}\mE^\H$
		\State $\vv_0\gets  \mQ_0^{-1}(\mA^\H\vybar_0 + \mB^\H\vybar_1)$
		\For{$K=1,2,\ldots$}
			\State Observe array sample batch $\vybar_{K+1}$
			\State $\mQ_{K} \gets  \mQ_0 - \mE\mU_{K-1}$
			\State $\mU_K \gets \mQ_K^{-1}\mE^\H$
			\State $\vv_K \gets \mQ_K^{-1}(\mA^\H\vybar_K + \mB^\H\vybar_{K+1}-\mE\vv_{K-1})$
			\State $\hat\valpha_{K+1} \gets (\mA^\H\mA+\delta\mId-\mE\mU_K)^{-1}\left(\mA^\H\vybar_{K+1} - \mE\vv_K\right)$
			\State $\hat\valpha_k\gets \mQ_k^{'-1}(\vw_k'-\mE_k\vv_k)$
			\For{$\ell = K,K-1,\ldots,0$}
				\State $\widehat\valpha_\ell \gets \vv_\ell - \mU_\ell\widehat\valpha_{\ell+1}$
			\EndFor
		\EndFor
	\end{algorithmic}
	\caption{\small\sl Streaming algorithm for solving \eqref{eq:lsK}.}
	\label{alg:streaming}
\end{algorithm}

With the algorithm defined, its is straightforward to gauge its computational complexity.  At each step, we have to compute $\mE\mU_{K-1}$ at a cost of $O(D^2MN)$, invert $\mQ_K$ at a cost of $O(D^3)$,  and compute $\mQ_K^{-1}\mE^\H$ at a cost of $O(D^3)$.  With a finite buffer, the backtracking consists of small number of vector-matrix multiplies for a cost pf $O(DMN)$.  As $D\ll MN$, the total computational cost is $O(D^2MN)$ per sample batch.

Note, however, that all of the matrices above are independent of the data and can be precomputed.  With this precomputation done, each update essentially consists of a small number of matrix-vector multiplies at a cost of $O(DMN)$.  Moreover, the matrices $\mQ_K$ and $\mU_K$ converge to fixed values relatively quickly, so the cost of these precomputations in manageable.

\subsection{Packet Merging}
As noted in Section~\ref{sec:biasvar} the variance term \eqref{eq:randomarraygain} and consequently the array gain are proportional to $D_N(\theta)/MN$. The term $2 \Omega T_1(\theta) + L-1$ prevents us from achieving the ideal array gain except in the limit as $N \to \infty$. Nevertheless, we can still improve array gain by making $N$ as large as possible e.g.\ by increasing the batch size. Doing this directly can be computationally difficult, and requires a longer buffer and consequently longer latency. An alternative approach is to merge packets as they exit the buffer.

To see why this works, let $\setS_k$ be the subspace spanned by $\{\psi_{k,d}(t)\}_{d=1}^D$. The union of $B'$ adjacent packet subspaces $\setS = \bigcup_{d=1}^{B'}\setS_{k-d}$ is by design spanned by the $DB'$ basis functions. However, we can approximate the same subspace by directly defining a subspace $\setS'$ that represents the packet over the support interval $ \bigcup_{d=1}^{B'}\setT_{k-d}$. The key detail is that $\setS' \subsetsim \setS$ while $\dim(\setS')\sim D_{NB'}(\theta)$. Since $L$ grows slowly with $N$ and we remove an additional and unnecessary $(B'-1)\cdot (2 \Omega T_1(\theta)-1)$ degrees of freedom, the dimension of $\setS'$ may be far smaller than $\setS$. In turn, the array gain associated with an approximation is $\setS'$ is greater than its nominal representation in $\setS$.

We leverage this fact by generating a mapping that takes the estimated signal, which lies in $\setS$, and projects it into $\setS'$. We apply this mapping to each set of $B'$ packets as they exit the buffer (e.g.\ when their coefficients are no longer being updated). A qualitative visualization of what this looks like is shown in Figure~\ref{fig:streaming_ls}(a) where we see the envelope of 3 packets merged into 1. Additionally, this update can be done efficently through a low-rank update as described in Appendix~\ref{apdx:lr_update}. 

\subsection{Numerical Experiments}
\begin{figure}[h]
	\centering
	\begin{tabular}{cc}
		\includegraphics[width=.45\textwidth]{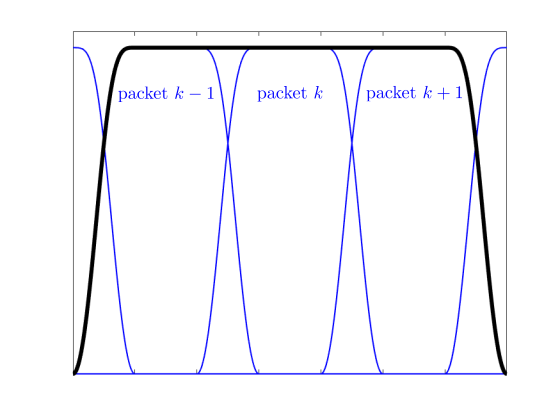}  & \includegraphics[width=.45\textwidth]{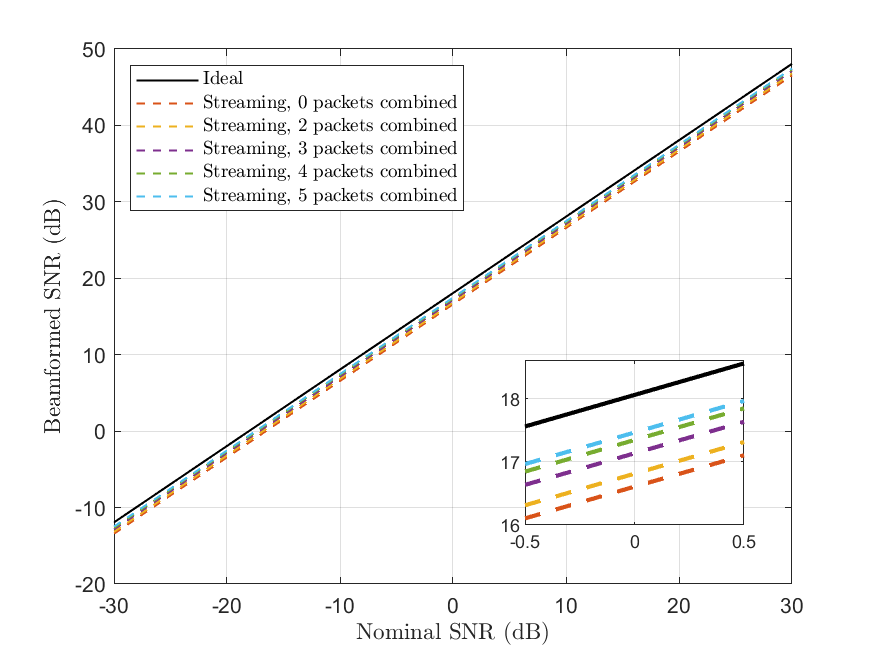}\\
		(a) Packet merging & (b) ULA
	\end{tabular}
	\caption{\small\sl (a) packet merging takes a predetermined number of packets that have exited the buffer (e.g.\ their coefficients are no longer being updated), and merges the individual packet subspaces which improves array gain. (b) For a $M=2^6$ ULA and a packet size of $N=2^5$ we reconstruct 120 batches (split into 120 packets) and calculate the beamformed SNR across the packets that have exited the buffer. It can be seen that merging a small number of packets drives the array gain closer to the ideal value.}
	\label{fig:streaming_ls}
\end{figure}

As a demonstration of the effectiveness of both the streaming least squares approach and the packet merging we again return to the ULA scenario we have examined throughout the paper. All parameters are the same as the previous experiments, except now we observe batches of $N=2^5$ samples off the array and reconstruct the signal in a streaming manner. The buffer size is set to $B=5$ and reconstruct 120 packets using Algorithm~\ref{alg:streaming}. We note that inverting the system directly with this number of packets would be computationally intractable for standard computing devices. The beamformed SNR is calculated only across the portion of the signal that had access to a full buffer, e.g.\ we neglect the first and last 5 packets in the calculation. Additionally we examine how merging a varying number of packets effects performance. Results were averaged over 50 trials from randomly generated signals as in our previous experiments. The results in Figure~\ref{fig:streaming_ls}(b) demonstrate that without any type of merger the streaming result performs as well as the non-streaming (e.g.\ single batch) result. Furthermore, leveraging our computationally efficient packet merging paradigm we can substantially improve performance, with a merger of 5 packets driving the beamformed SNR close to the ideal gain.

\section{Dimensionality reduction}
\label{sec:dimreduce}
In Section~\ref{sec:subbuls}, Section~\ref{sec:compcomplex}, and Section~\ref{sec:slepianmvdr} we implicitly introduced the idea of dimensionality reduction at the sensor via encodings. Instead of reading samples directly off the sensor elements and fitting our Slepian subspace model we observe the samples through linear measurements and fit our model accordingly. The previously discussed examples are a few specific cases, but as we will see in this section the concept can be easily generalized to measurements taken across space or space and time. This provides a means of substantially reducing the array readout, which can offer several benefits in terms of practical implementation.

\subsection{Beamforming from linear encodings using least squares}
\label{sec:bulsrd}

We model the dimensionality reduction operation as a $P \times MN$ matrix $\mPhi$ (with $ D_N(\theta)\leq P \ll MN$) that is applied to the collection of snapshots $\underline\vy$. We denote the linear measurements taken from the array by the $P \times 1$ vector
\begin{align}
    \label{eq:w_encode}
    \underline\vw = \mPhi\underline\vy.
\end{align}
We can estimate the coefficients $\valpha$ directly from \eqref{eq:w_encode} through a slightly modified version of \eqref{eq:lsconv} 
\begin{equation}
	\label{eq:lsconv_rd}
	\minimize_{\valpha\in\C^{D_N(\theta)}}~\frac{1}{2}\|\underline\vw - \mPhi\mA(\theta)\valpha\|_2^2 + \delta\|\valpha\|_2^2.
\end{equation}
Of course, the ability to estimate $\valpha$ and its effectiveness is dependent on the conditioning of the $P \times D_N(\theta)$ composite matrix $\mPsi = \mPhi\mA(\theta)$. The regularization parameter $\delta$ helps to alleviate some of the concerns on the conditioning of $\mPsi$. However, for a properly chosen $\mPhi$ we can expect $\mPsi$ to be well conditioned and can estimate $\valpha$ close to or as well as if we had access to $\vybar$ directly.

As alluded to above, a reasonable choice in $\mPhi$ can actually be relatively arbitrary. To make this point more clear consider the case where the $\mPhi$ has i.i.d.\ entries drawn from a Gaussian or subgaussian distribution. If $P = O(D_N(\theta))$, then as a direct consequence of \cite[Lemma 1.1]{davenport2013si} with high probability
\begin{align*}
    c_3 \cdot \sigma_{\text{min}}(\mA(\theta))\leq \sigma_i(\mPsi) \leq c_4 \cdot \sigma_{\text{max}}(\mA(\theta)), ~ \text{for} ~ i = 1,2,\dots,D_N(\theta),
\end{align*}
where the constants $c_3 < 1$ and $c_4>1$ can be driven close to $1$ by increasing $P$. What this tells us is that even when taking a set of random measurements $\mPsi$ will be conditioned comparably to $\mA(\theta)$ as long as we take enough measurements. Therefore estimating $\valpha$ through application of the pseudo-inverse of $\mPsi$ will have a comparable bias to estimation from direct snapshots. On the other hand, random measurements will induce more variance (and consequently reduce array gain) as will be apparent in our numerical experiments. This can be mitigated by increasing the number of measurements, but they will always generally underperform in comparison to more structured measurements.

Although dimensionality reduction with random $\mPhi$ illustrates the flexibility of our choice in measurements, it is not how one would generally choose to design $\mPhi$. It is computationally inefficient to apply a random matrix, and $P$ exceeds $D_N(\theta)$ meaning we are sub-optimally reducing dimension. In the next sections we will examine two different measurement designs that more directly leverage our Slepian subspace model. The first examines how to encode snapshot-by-snapshot, taking measurements across spatial array elements independent of the temporal sampling instance. The second encodes across space and time by measuring and combining all snapshots simultaneously.

\subsection{Spatial encoding}
\label{sec:dimredspat}
Spatial encodings, named such because they linearly combine across array elements, handle the array output snapshot-by-snapshot. In this paradigm the measurement matrix and subsequent measurements are structured as 
\begin{align*}
    \mPhi = 
    \begin{bmatrix}
        \mPhi_1 & \mtx{0} & \cdots & \mtx{0}\\
        \mtx{0}  & \mPhi_2 & \ddots & \vdots\\
        \vdots & \ddots & \ddots & \mtx{0} \\
        \mtx{0} & \cdots & \mtx{0} & \mPhi_N
    \end{bmatrix},~
    \underline\vw = 
    \begin{bmatrix}
        \mPhi_1 \vy[1]\\
        \mPhi_2 \vy[2] & \\
        \vdots \\
        \mPhi_N \vy[N]
    \end{bmatrix},
\end{align*}
where each submatrix $\mPhi_n$ is $P_n \times M$ such that $P = \sum_{n}P_n$. Hence each non-zero submatrix of $\mPhi$ only interacts with a single snapshot. 

We have actually already seen two instances of this kind of measurement with the first being the subarray beamforming method discussed in Section~\ref{sec:subbuls}. To see how this translates to Slepian beamforming we first vectorize the operation in \eqref{eq:ysub} by defining a $M \times 1$ vector $\vphi_{m'}$ where $\vphi_{m'}[i] = w_{i}$ for $i \in \mathcal{M}_{m'}$ and $\vphi_{m'}[i] = 0$ for $i \notin \mathcal{M}_{m'}$. A sub-beam from a snapshot $\vy[n]$ is formed by simply taking $\vphi_{m'}^H \vy[n]$. Collecting these weight vectors into a $P_n \times M$ matrix $\mPhi_{n}^H = [\vphi_{1},\vphi_{2},\dots,\vphi_{M'}]$ where $P_n = M'$ all sub-beams from a snapshot are formed simultaneously by taking $\mPhi_n \vy[n]$. Hence it directly maps into our spatial encoding scheme and we can accurately estimate $\valpha$ from $P = NM'$ measurements leveraging \eqref{eq:lsconv_rd} as previously demonstrated.

The second instance we have seen was the efficient least squares approach discussed in Section~\ref{sec:compcomplex}. As was noted, each snapshot lies in its own Slepian space spanned by the $M\times D_1(\theta)$ orthonormal matrix $\mU$. The first step in the approach is forming encodings $\vbeta_n = \mU^H \vy[n]$ and is not too difficult to see that this is equivalent to setting $\mPhi_n = \mU^H$. As we have seen this yields good estimates of $\valpha$ from $P = N \cdot D_1(\theta)$ measurements.

Though in general each $\mPhi_n$ can differ, these two examples highlight that fixing measurements such that $\mPhi_1 = \mPhi_2 = \dots = \mPhi_N$ works well and is simpler to implement practically. In particular $\mPhi$ could be applied in the analog domain via a vector matrix multiply with fixed weights prior to sampling \cite{sharma2023af,rahman2024be}. Estimating the signal by solving \eqref{eq:lsconv_rd} could then be done digitally.

\subsection{Spatial temporal encoding}
Though spatial encoding alone can greatly reduce the dimensionality of the array output it does not leverage the full structure of our model. If instead we allow ourselves to encode across snapshots (e.g.\, we don't constrain $\mPhi$ to have a block-diagonal structure), then we can further collapse the dimension of $\vybar$ with \eqref{eq:w_encode}. We have seen operations like this previously in Section~\ref{sec:slepianmvdr} where setting $\mPhi = \mW$ reduced $\vybar$ from dimension $MN$ to $D_N(\theta)$ while nulling an interferer. If an interferer were not present we could simply set $\mPhi = \mA^{\dagger}(\theta)$ to extract the coefficients directly or set $\mPhi = \mA^{H}(\theta)$ and perform the remainder of the least squares step later. Each of these three $\mPhi$ are $D_N(\theta) \times MN$ and hence their application to $\vybar$ reduces the dimension of the array readout to $D_N(\theta)$. Since the subspace dimension is $D_N(\theta)$, these forms of measurement are essentially the best we can do in terms of dimensionality reduction under our given model.

The discussion in Section~\ref{sec:bulsrd} on random measurements tells us that we can estimate the signal from unstructured measurements. The main trade-off being we must take more measurements and there will be a higher variance. Though we likely would not utilize random measurements in practice it reaffirms the point that recovery is achievable for most reasonable choices in $\mPhi$. To this end, a family of ``hardware-friendly" measurements that take into account non-idealities in the measurement process are discussed in \cite{delude2022bro}. The conclusion is the same as ours here: with a modest increase in the number of measurements we can achieve good performance. However, the measurement types are far more structured and can be implemented in either purely spatial or spatial temporal encoding schemes.

\subsection{Numerical Experiments: Dimensionality reduction}

The results shown in Figure~\ref{fig:efficient_ls_alpha_diff} demonstrate how spatial Slepian encodings behave. Similarly Slepian spatial temporal encodings (e.g.,\ $\mPhi = \mA^{\dagger}(\theta)$) will produce similar results as Figure~\ref{fig:conv_bf_ag} for the respective choices in $L$. Furthermore, the spatial and spatial temporal encoding schemes discussed above $\mPsi$ will almost certainly possess full column rank, meaning it has a left inverse. As a result we do not expect the truncation bias or mismatch bias to change drastically. However, the variance, which in this case is given by $\sigma^2 \trace{\left (\mPsi^{\dagger} \mPhi\mPhi^H\mPsi^{\dagger H} \right )}$, may be substantially impacted by a particular choice in $\mPhi$. In order to reduce the level of redundency in our experiments, this section will simply examine the variance trace term as an indicator for how $\mPhi$ impacts array gain.

Our experiments return to the ULA and UPA examples we have used throughout the paper. For each respective array geometry we examine Slepian and i.i.d. Gaussian random $\mPhi$ in spatial and spatial temporal encoding configurations. We treat the Slepian spatial temporal encoding as a baseline since we expect it to be relatively agnostic to the number of measurements taken. It will also yield the same variance that having direct access to $\vybar$ will produce. For the other encoding schemes we vary the number of measurements and calculate the corresponding $\trace{\left (\mPsi^{\dagger} \mPhi\mPhi^H\mPsi^{\dagger H} \right )}$. For random measurements we average this term over 50 trials, drawing new random $\mPhi$ in each trial.

\begin{figure}[h]
	\centering
	\begin{tabular}{cc}
		\includegraphics[width=.45\textwidth]{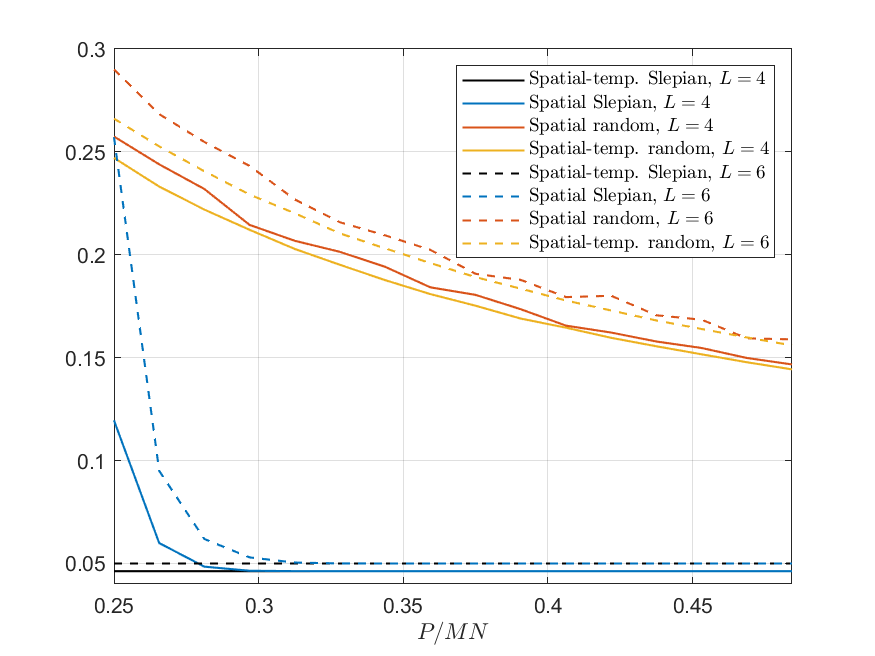}  & \includegraphics[width=.45\textwidth]{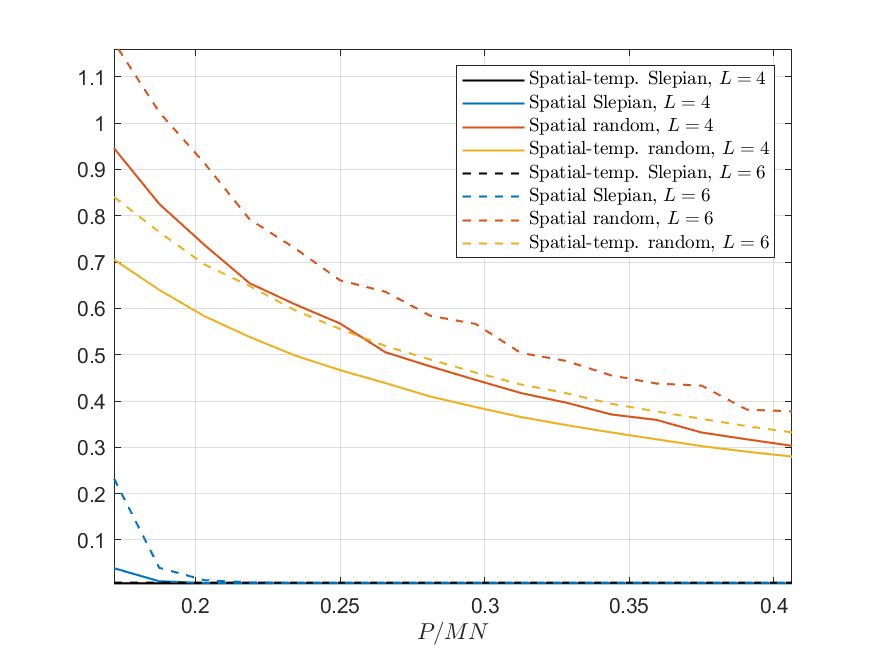}\\
		(a) 64 element ULA & (b) $32\times32$ element UPA
	\end{tabular}
	\caption{\small\sl Variance multiplier $\trace{\left (\mPsi^{\dagger} \mPhi\mPhi^H\mPsi^{\dagger H} \right )}$ of Slepian subspace beamforming for collections of $N=2^5$ snapshots when subject to dimensionality reduction prior to coefficient estimation for the (a) ULA and (b) UPA used in our previous experiments. We examine spatial and spatial-temporal encoding schemes while varying the number of measurements we take (e.g.\ the parameter $P$). The spatial-temporal Slepian encoding is simply the result of directly embedding in the Slepian space we approximate the signal with, representing the optimal result. }
	\label{fig:dim_reduction_variance}
\end{figure}

The results shown in Figure~\ref{fig:dim_reduction_variance} indicate that the Slepian spatial encodings perform as well as the complementary spatial temporal encodings as long as $P$ exceeds 650 and 450 for the ULA and UPA array case respectively. This means leveraging only spatial encodings we can beamform with effectively no loss to performance using $31.7\%$ and $1.3\%$ of the number of samples in $\vybar$ for the ULA and UPA cases respectively. Of course, this percentage can be reduced even further if we allow for temporal encoding as well. The spatial temporal Slepian encodings perform identically to having full access to $\vybar$ but only require using $1.8\%$ and $0.15\%$ the number of samples for the ULA and UPA respectively. The random measurements induce a substanially higher variance than their Slepian counterparts. Increasing $P$ we can drive the variance down since increasing the number of measurements will have noise averaging affects and the error is largely noise dominant. Ultimately, by taking enough measurements, unstructured $\mPhi$ can perform reasonably well.


\subsection{Streaming reconstruction with dimensionality reduction}

\begin{figure}[h]
	\centering
	\begin{tabular}{cc}
		\includegraphics[width=.45\textwidth]{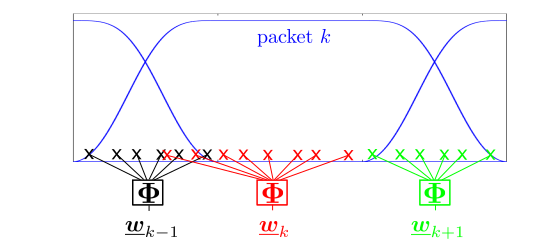}  & \includegraphics[width=.45\textwidth]{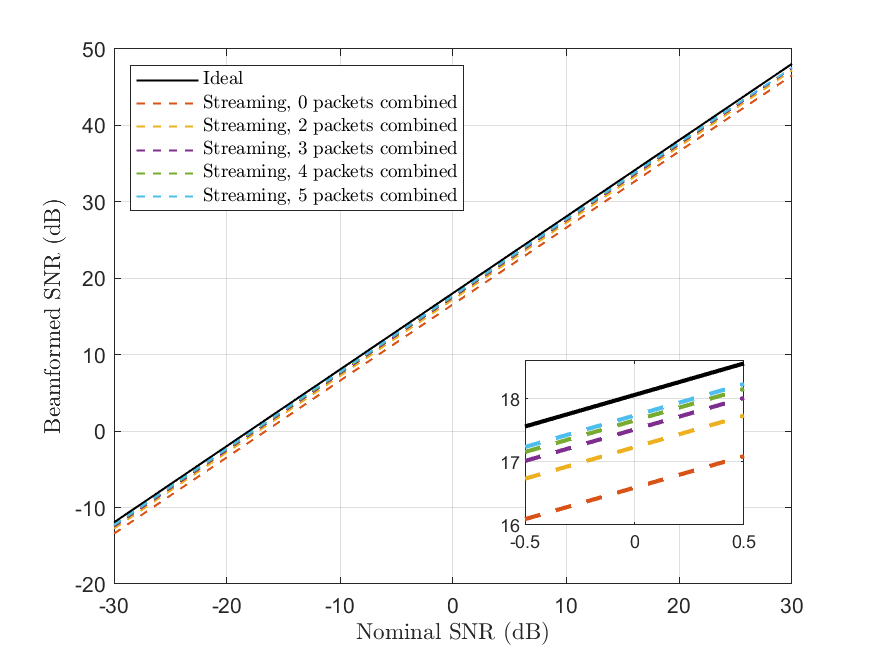}\\
		(a) Packet merging & (b) ULA
	\end{tabular}
	\caption{\small\sl (a) Linear measurements are acquired across each packet by applying a suitable $\mPhi$ matrix. For things to remain the same algorithmically we require that temporal extent of each measurements only interacts with at most 2 packets at any given time. (b) For a ULA with a batch size of $N=2^6$ we use a slightly modified version of Algorithm~\ref{alg:streaming} and spatial Slepian measurements to reconstruct the signal. We reconstruct $K=120$ blocks (split into 120 packets), and compute the SNR amongst all packets that had access to a full buffer.}
	\label{fig:streaming_ls_meas}
\end{figure}

The streaming reconstruction method described in Section~\ref{sec:streaming} can be easily extended to streaming reconstruction from dimensionality reducing measurments. All that we require is that the temporal extent of the measurements do not interact or ``touch" two packets at the same time. A qualitative depiction of this is shown in Figure~\ref{fig:streaming_ls_meas}(a) where we can see all the measurements are linearly combining samples only residing in at most two packets. Since the temporal support of the measurements bleed past the overlap region, this has the affect of increasing the overlap between successive packets. As such, it may be required to increase the batch size $N$ for this constraint to hold. Upon satisfying this requirement, a careful (but straightforward) re-partitioning of each $\setT_k$ to account for the ``new" overlap regions is performed. Subsequently we form matrices $\mA$ and $\mB$ in the exact same manner as before and modify \eqref{eq:lsK} to
\begin{equation}
	\label{eq:lsK_meas}
	\minimize_{\{\valpha_0,\ldots,\valpha_K\}}~\sum_{k=0}^{K} \left\|\mPhi\mA\valpha_k +\mPhi\mB\valpha_{k-1} - \underline{\vw}_k\right\|_2^2 + \delta\sum_{k=0}^K\|\valpha_k\|_2^2.
\end{equation}
such that we are operating directly on measurements. If we directly substitute $\mPhi\mA$, $\mPhi\mB$, and $\underline{\vw}_k $ for $\mA$, $\mB$, and  $\vybar_k$ in Algorithm~\ref{alg:streaming} nothing changes algorithmically. What does change is the computational complexity, which can be reduced to $O(D^2)$ per sample batch when using spatial temporal Slepian measurements since $P = D \ll MN$. The previously described packet combining operations remain the same as well.

As a brief demonstration of the performance of the dimensionality reduced streaming reconstruction algorithm we return to the ULA scenario we have used throughout this paper. We Set $\mPhi$ to be the spatial Slepian encoding scheme discussed in Section~\ref{sec:dimredspat}. In order to insure the measurements do not interact with more than two packets at a time we increase the batch size to $N=2^6$, and again the buffer size is set $B = 5$. We reconstruct 120 packets using the slightly modified version of Algorithm~\ref{alg:streaming} and calculate the beamformed SNR across all packets that had access to a full buffer (e.g.\ we neglect the first and last 5 packets). As in the previous experiments we average results over 50 trials with randomly generated signals. The results shown in Figure~\ref{fig:streaming_ls_meas}(b) show essentially identical trends to our results using full array snapshots, albeit with a better array gain due to the increased batch size. As before we see that packet combining as packets exit the buffer greatly increases array gain.

\section{Conclusion}
\label{sec:conclusion}
We have introduced a new method of broadband beamforming leveraging a Slepian subspace model, circumventing the need for direct filtering. Using either single or multiple snapshots we estimate the signal's expansion coefficients using least squares, and subsequently use these coefficients to approximate the Nyquist rate samples. This process can take place on individual blocks of samples, or on multiple blocks in a streaming manner. We have demonstrated how this methodology can be naturally extended to adaptive beamforming, allowing for efficient and effective interferer cancellation. Finally, in essentially all tested scenarios our proposed method outperformed more conventional approaches while also carrying a comparable computational cost.


\newpage 

\bibliographystyle{IEEEtran}
\bibliography{IEEEabrv.bib,broadbeamrefs}

\appendix
\section{Frequency invariant beamforming low frequency attenuation}
\label{apx:fibf}

Frequency invariant beamformer design leverages a carefully chosen dilation and truncation of a desired array response to make performance uniform across all frequencies within a desired band\cite{ward1995th}. One of the main strengths of this non-linear process is that it defines a way to map narrowband beam designs directly into a broadband regime. Therefore we can leverage all of our knowledge and intuition behind narrowband beamforming (which is inherently simpler than broadband beamforming) to perform broadband beamforming. 

\begin{figure}[h]
	\centering
    \begin{tabular}{cc}
		\includegraphics[width=.45\textwidth]{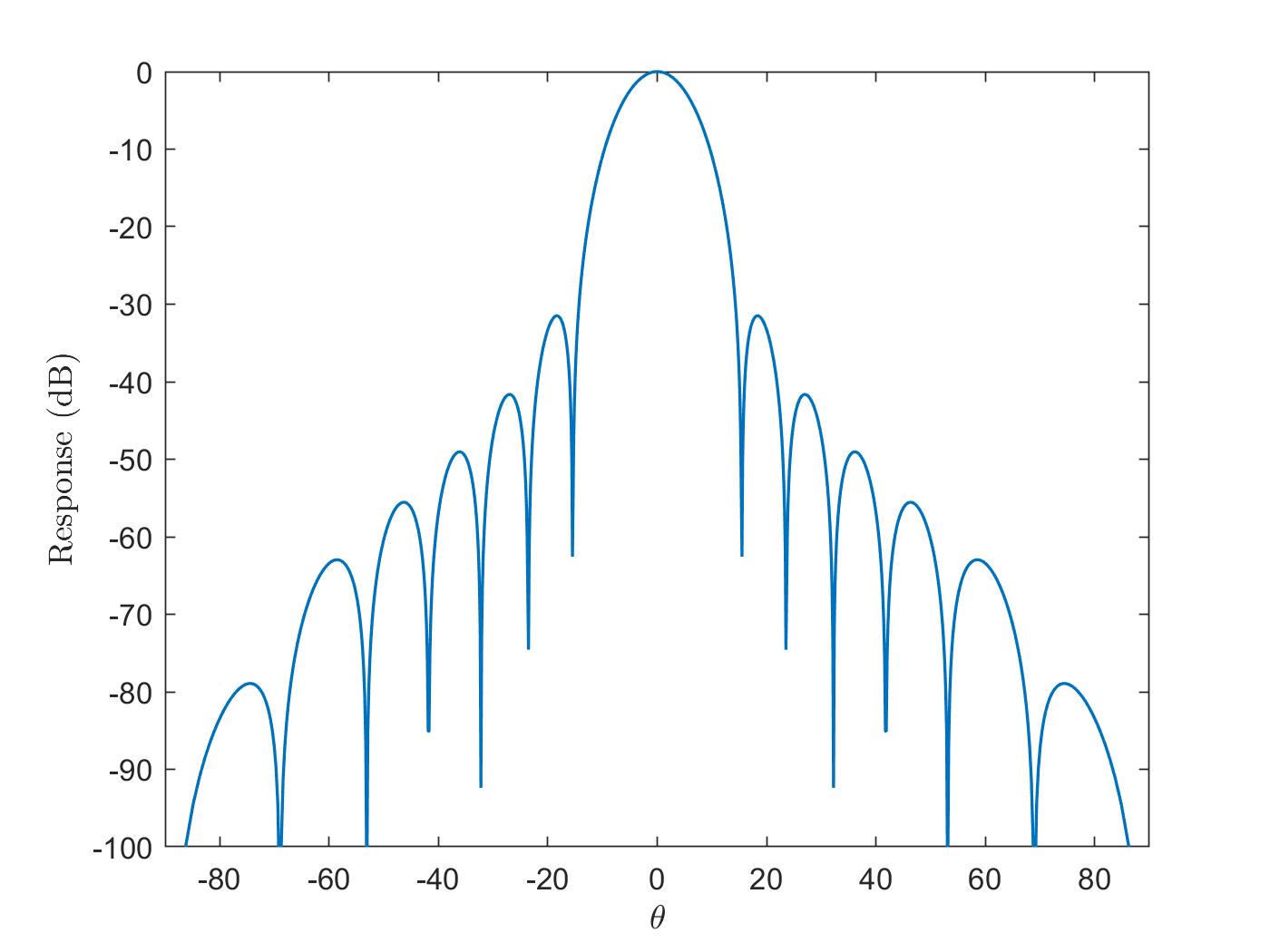}  & \includegraphics[width=.45\textwidth]{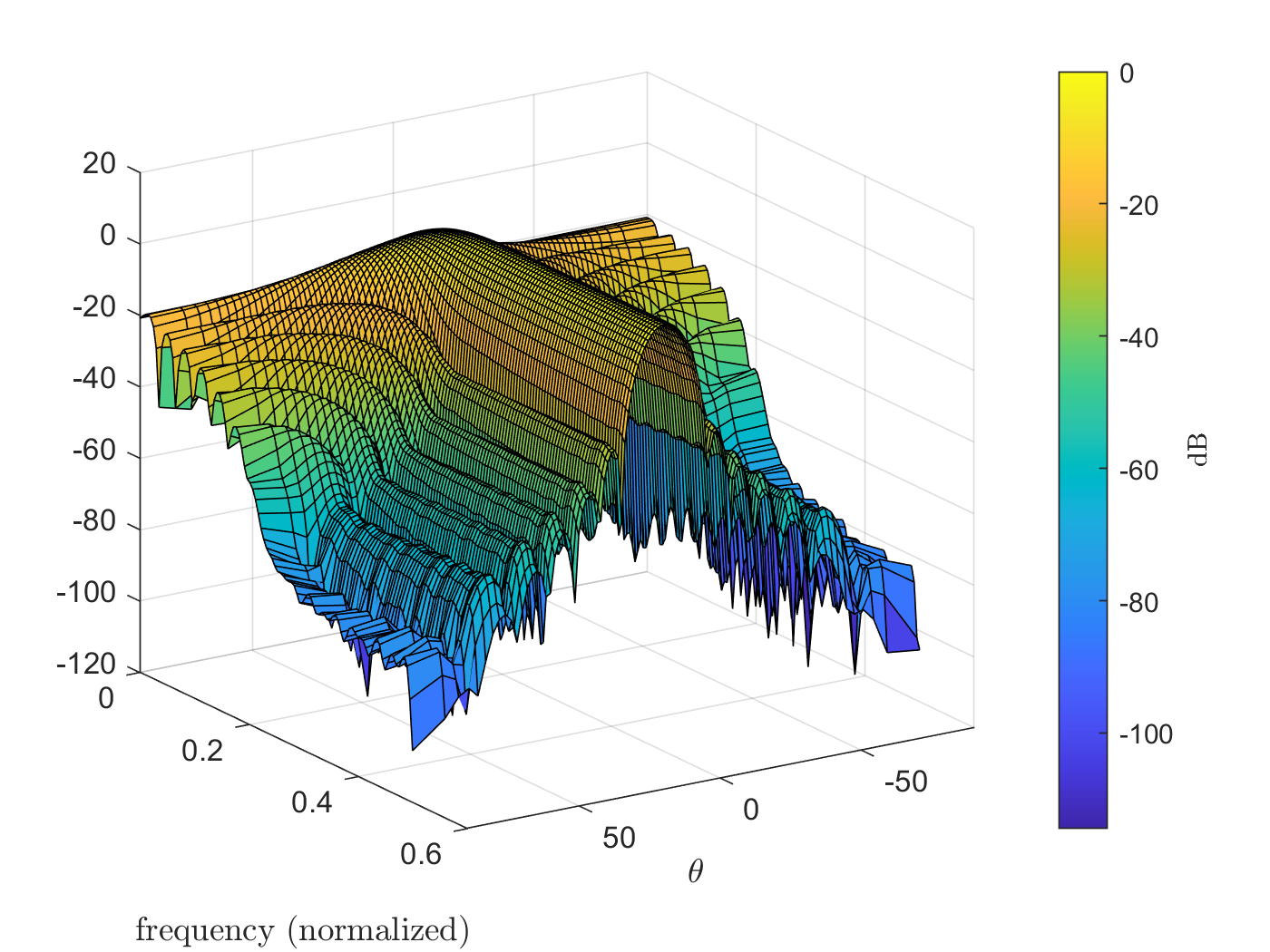}\\
		(a) Narrowband response & (b) Mapped Frequency invarient response
	\end{tabular}
	\caption{\small\sl (a) Narrowband beampattern generated from a 64-element ULA using standard weight and sum with Hann windowing. (b) The response of the  resultant frequency invariant beamformer's filter bank to sinusoidal signals at a variety of incident angles and frequencies. Though there is unity gain at a fixed angle over a wide variety of frequencies, the attenuation of the lower frequencies leads to a high level of distortion when the baseband signal has substantial energy in the lower bands.}
	\label{fig:fib_beampattern}
\end{figure}

The most straightforward approach to designing a frequency invariant beam is to first choose a desired response such as the one shown in Figure~\ref{fig:fib_beampattern}(a) which is simply a standard Hann windowed narrowband beam for a 64-element ULA. We apply the appropriate frequency domain mapping at a descrete set of evaluation points as described in \cite[Ch. 5]{liu2010wi} and then map to the time-domain via a DFT and truncation. The result is a filter bank that produces a frequency invariant beampattern (See Figure~\ref{fig:fib_beampattern}(b)). A caveat of this is process is that due to the conic support window of the desired frequency response the lower frequencies have a lower sampling density than higher frequencies. Hence the response of the beam to lower frequencies cannot be modeled as accurately \cite{liu2008de,liu2010wi}. As a result there is substantial attenuation of lower frequency signal components as shown in Figure~\ref{fig:fib_beampattern}(b).
In applications where baseband processing is not a necessity this is not particularly prohibitive. However, for most RF applications some form of down-conversion is necessary and hence the frequency invariant beamformers designed in the described manner will induce substantial distortion. 

\begin{figure}[h]
	\centering
    \begin{tabular}{cc}
		\includegraphics[width=.45\textwidth]{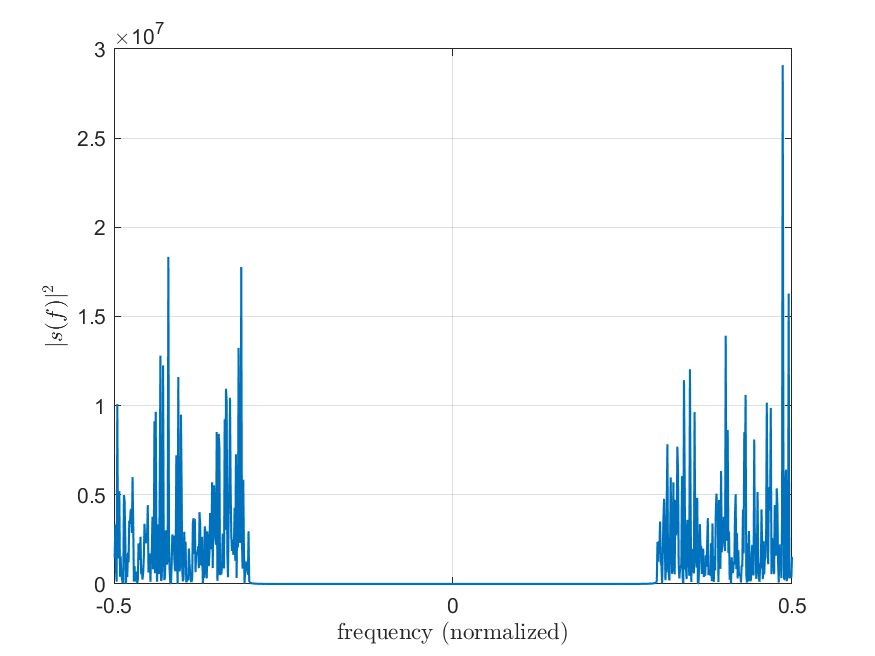}  & \includegraphics[width=.45\textwidth]{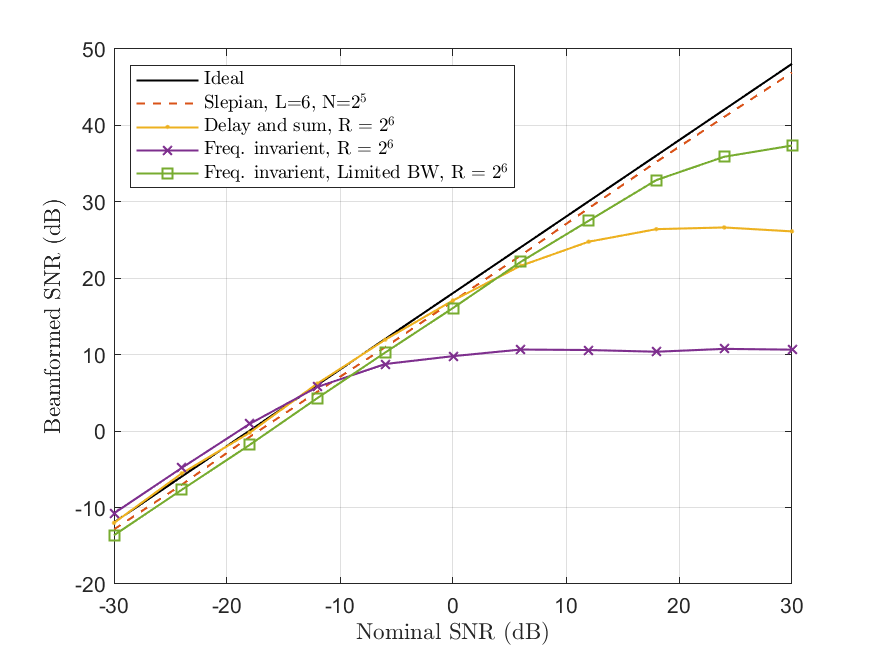}\\
		(a) Signal with frequencies only in upper band & (b) 64 element ULA
	\end{tabular}
	\caption{\small\sl (a) A signal with the same bandwidth as previous experiments, but with active frequencies limited to the upper portion of the occupied band. (b) Nominal vs beamformed SNR for 64-element ULA using Slepian subspace beamforming, delay and sum via fractional delay filtering beamforming, frequency invariant beamforming for a fully occupied band, and frequency invariant beamforming for signals with a partially occupied band.}
	\label{fig:fib_comp}
\end{figure}

To make this point more clear, consider a sampled bandlimited signal where the active frequency components are limited to the upper portion of the band as shown Figure~\ref{fig:fib_comp}(a). If we run a complementary set of experiments to Section~\ref{sec:conventionalexperiments} on signals of this type the resulting trend for a frequency invariant beamformer is shown in Figure~\ref{fig:fib_comp}(b). We see a tremendous improvement over the original frequency invariant beamformer's performance and although the Slepian subspace beamformer still offers the most consistent performance the performance gap is greatly narrowed. 

One manner in which we can ``force" the signal to have a structure like Figure~\ref{fig:fib_comp}(a) is to oversample, which naturally leaves a portion of the spectrum unoccupied. After proper modulation the signal could then be beamformed using frequency invariant beamformers without suffering from low frequency attenuation. As a brief comment, this is counter to the manner in which fractional delay filter performance can be improved. In general, fractional delay filters incur the largest portion of their phase and amplitude distortion at the highest frequencies. Oversampling improves performance by making the signal more concentrated in the lower frequencies, since any frequency component past the Nyquist frequency will be 0.  
\section{Approximate covariance matrix inversion}
\label{apx:covaprx}

If $\mC$ is an invertible matrix and $\mU$ and $\mtx{I}$ are conformal, then it is a fact that $\mU\mC\mU^H + \sigma^2 \mtx{I}$ has the inverse
\begin{align*}
    (\mU\mC\mU^H + \sigma^2 \mtx{I})^{-1} = \frac{1}{\sigma^2}\mtx{I} - \frac{1}{\sigma^4}\mU\left (\mC^{-1} + \frac{1}{\sigma^2} \mU^H\mU\right)^{-1}\mU^H
\end{align*}
as given by the Woodbury inversion lemma \cite{woodbury1950in}. When $\mU\in \mathbb{C}^{M\times L}$ and $\mC\in \mathbb{C}^{L\times L}$ with $L\ll M$, leveraging this lemma to invert a large matrix can yield substantial computational savings. However, in order to leverage this we must have a low rank factorization in hand. 

\begin{figure}[h]
	\centering
	\begin{tabular}{cc}
		\includegraphics[width=.45\textwidth]{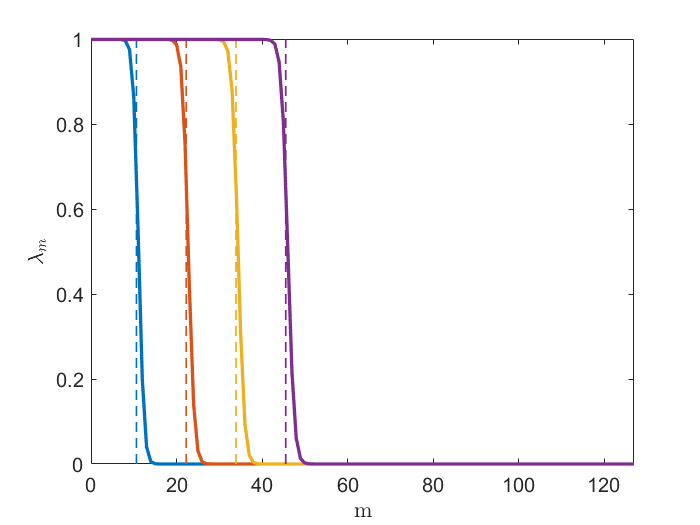}  & \includegraphics[width=.45\textwidth]{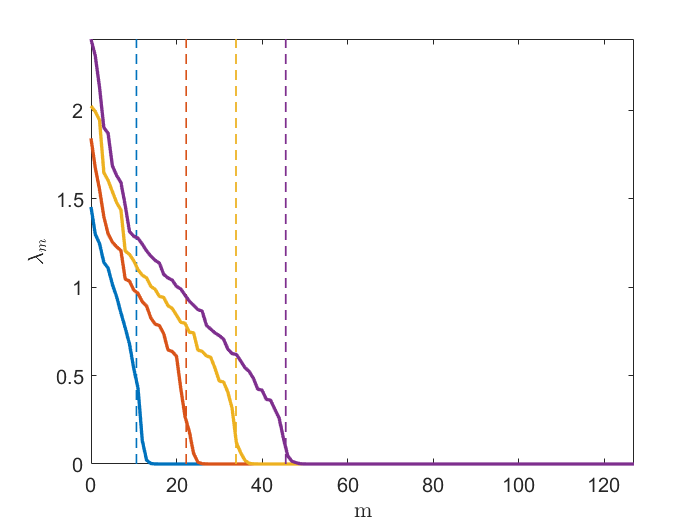}\\
		(a) Uniform samples & (b) Non-uniform samples
	\end{tabular}
	\caption{\small\sl Eigenvalue concentration of covariance matrices generated by sampling a Gaussian random process with flat power spectral density (a) uniformly and (b) non-uniformly over a fixed time interval and several different bandwidths. The dashed lines indicate the time-bandwidth produce associated with each process.}
	\label{fig:prolate_eigs}
\end{figure}

Luckily, the signal and interferer covariance matrices are \emph{numerically} low-rank.\footnote{Meaning that the vast majority of their eigenvalues are close to 0, but not actually 0.} To help visualize this, consider a Gaussian random process with flat power spectral density (as was discussed in detail in Section~\ref{sec:slepian}). If we sample this signal $M$ times over a fixed interval the resultant eigenspectrums of the covariance matrices shown in Figure~\ref{fig:prolate_eigs} exhibit sharp decay as eigenvalue indices exceed the time-bandwidth product. Furthermore, this property holds for any bandwidth and whether the samples are taken uniformly or non-uniformly. 

With this favorable property in hand we can approximate signal and interferer covariance matrix by simply truncating the eigenvalues such that
\begin{align*}
    \mR_S + \mR_I &= \mV\mLambda\mV^H + \mW\mS\mW^H\\
    &\approx \mV_K\mLambda_K\mV_K^H + \mW_{K'}\mS_{K'}\mW_{K'}^H\\
    & = [\mV_K, \mW_K]
    \begin{bmatrix}
    \mLambda_K & \mtx{0}\\
    \mtx{0} & \mS_{K'}
    \end{bmatrix}
    [\mV_K, \mW_K]^H\\
    & = \mU\mC\mU^H.
\end{align*}
In this case $\mU \in \mathbb{C}^{M\times (K+ K')}$, and based on Section~\ref{sec:slepian} both $K$ and $K'$ grow very slowly in comparison to $M$. Coupling this with the fact the noise covariance matrix is simply a scaled identity matrix we can efficiently invert the low-rank approximated covariance matrix using the previously mention inversion lemma.

One issue with this is that the forming a low-rank factorization is as difficult computationally as inverting the the matrix. However, we know that the column space of the signal and intereferer covariance matrix is spanned by the Slepian basis vectors, which we can calculate directly. Furthermore, using the same methedology, we can calculate the eigenvalues directly as well. This is a straightforward process in the uniformly samples case, but translating it to the non-uniform case requires slightly more care.

A simple way to deal with the non-uniform case is to utilize the fact that uniform samples of the Slepian basis vectors are easy to generate, smooth, and are samples of the underlying continuous Slepian functions. Let $\mS_u$ be the uniformly sampled Slepian basis vectors and $\widehat\mR_u \approx \mS_u \mLambda_u \mS_u^H$ be a low rank approximation to a uniformly sampled Gaussian random process with flat power spectral density. Suppose we want to determine the covariance matrix $\mR_{nu}$ for non-uniform samples taken over the same interval. Let $\mH$ be an interpolation matrix,\footnote{In general interpolation of a smooth function can be done very efficiently.} then we can interpolate the $\mS_u$ onto the non-uniform sample points to form $\mS_{nu} = \mH \mS_u$. The $\mS_{nu}$ will span the column space of $\mR_{nu}$ but lose their orthogonality. For additional computational savings we can perform the simple factorization $\mS_{nu} = \mU_{nu} \mSigma_{nu} \mV_{nu}^H$ such that
\begin{align*}
    \widehat\mR_{nu} &= \mU_{nu} \mSigma_{nu} \mV_{nu}^H \mLambda_u\mV_{nu}\mSigma_{nu}\mU_{nu}^H\\
    & = \mU_{nu} \mC_{nu} \mU_{nu}^H.
\end{align*}
This is not strictly necessary but saves a step in the application of the inversion lemma.

In summary, the fact that the signal and interferer covariance matrix is numerically low-rank provides an avenue for efficiently inverting the total covariance matrix. Using the fact the column space is spanned by the Slepian functions we can either directly calculate the low-rank factorization or interpolate to find it. Hence we do not need to directly invert or directly factorize any of the covariance matrices allowing us to test very large arrays.
\section{Packet combining via a low rank update}
\label{apdx:lr_update}

As discussed in Section~\ref{sec:streaming} we can improve array gain by successively combining packets as they exit the main buffer. The $k$th packet lies in the subspace $\setS_k$ with $\dim{(\setS_k)} = D_N(\theta) = \lceil 2\Omega T_N(\theta)\rceil + L$. If we denote $\setS = \bigcup_{k=1}^K \setS_k$ be the union of $K$ packet subspaces, then $\dim{(\setS)} = K \cdot D_N(\theta) = K\lceil 2\Omega T_N(\theta)\rceil + KL$. Building on our previous discussion, to within the same level of accuracy $s(t)$ approximately lies in a subspace $\setS'$ over the interval $T_{KN}(\theta)$. This subspace has $\dim{(\setS')} = D_{KN}(\theta) = \lceil 2\Omega T_{KN}(\theta)\rceil + L'$ where $\lceil 2\Omega T_{KN}(\theta)\rceil \leq K\lceil 2\Omega T_{N}(\theta)\rceil$ and since the truncation parameter grows slowly $L' \leq KL$. Hence the dimension of this subspace is possibly far smaller than $\setS$, and since a smaller subspace equates to greater array gain it is preferable to project our signal into this space.

If $\vy$ contains $D$ sample estimates of $s(t)$ formed by solving \eqref{eq:lsK} over the interval $T_{KN}(\theta)$, then $\vy \in \setS$. Let $\mP_{\setS'}$ denote the orthogonal projection operator onto $\setS'$, then we can compute the projection $\vy' = \mP_{\setS'}\vy$ directly with $D^2$ computations. If $D\gg D_{KN}(\theta)$ we can improve upon this by defining a $N \times D_{KN}(\theta)$ matrix $\mV_{\setS'}$ such that $\mV_{\setS'}^H\mV_{\setS'} = \mtx{I}$ and $\setS' = \Span(\mV_{\setS'})$. Letting $\mP_{\setS'}=\mV_{\setS'}\mV_{\setS'}^H$ we can apply the projection in $2\cdot D \cdot D_{KN}(\theta)$ computations. Howevever, in general we likely only care about producing the Nyquist rate samples and therefore $D$ will be close to $D_{KN}(\theta)$. Therefore the computational saving using this method are limited.

Real computational savings can be realized if we leverage the signals structure more thoroughly. To begin we note that $\setS' \subsetsim \setS$ e.g.\ $\setS'$ is approximately contained in the range of $\setS$. This hypothesis is a direct result of our discussions on approximating a signal over overlapping intervals in Section~\ref{sec:streaming}, and is only approximate due to the finite truncation of the basis expansion. Similar to before, define the $N \times \dim(\setS)$ matrix $\mV_{\setS}$ such that $\mV_{\setS}^H\mV_{\setS} = \mtx{I}$, and define the orthogonal projection operator onto $\setS$ as $\mP_{\setS} = \mV_{\setS}\mV_{\setS}^H$. Additionally, let $\mP_{\setS^{\perp}}$ be the projection operator onto the orthogonal compliment of $\mP_{\setS}$. With these definitions in hand we want to examine the difference term $\vy - \mP_{\setS'}\vy$. Reiterating the fact that $\vy \in \setS$ we have 
\begin{align*}
    \vy - \mP_{\setS'}\vy &= \mP_{\setS}\vy - \mP_{\setS'}\mP_{\setS}\vy\\
    &= \mP_{\setS}\vy - \mP_{\setS}\mP_{\setS'}\mP_{\setS}\vy - \mP_{\setS^{\perp}}\mP_{\setS'}\mP_{\setS}\vy\\
    & = \mV_{\setS}\mV_{\setS}^H\vy - \mV_{\setS} \underbrace{\mV_{\setS}^H\mV_{\setS'}\mV_{\setS'}^H\mV_{\setS}}_{\mG}\mV_{\setS}^H\vy - \mP_{\setS^{\perp}}\mP_{\setS'}\mP_{\setS}\vy\\
    & = \mV_{\setS} (\mtx{I} - \mG) \mV_{\setS}^H\vy - \underbrace{\mP_{\setS^{\perp}}\mP_{\setS'}\mP_{\setS}\vy}_{\text{misalignment-term}}.
\end{align*}
where the misalignment term is due to the fact $\setS'$ is only approximately contained in $\setS$. If it is in fact fully contained, then this term is 0. The matrix $\mG$ is $\dim(\setS) \times \dim(\setS)$, but $\rank(\mG) = \dim(\setS')$. It is symmetric-positive semidefinite and therefore has the eigendecomposition $\mG = \mW\mLambda\mW^H$ with $\mW\mW^H = \mW^H\mW = \mtx{I}$. Furthermore $\mLambda$ has structure 
\begin{align*}
    \mLambda = 
    \begin{bmatrix}
    \mC & \mtx{0}\\
    \mtx{0} & \mtx{0}
    \end{bmatrix},
\end{align*}
where $\mC$ is a diagonal matrix. The diagonal matrix $\mC$ contains the eigenvalues $\mC[i,i] = \cos^2\theta_i$ for $i = 1,\dots, \dim(\setS)$ where $\{ \theta_i \}_{i=1}^{\dim{(\setS')}}$ are the principle angles between the subspaces $\setS$ and $\setS'$ \cite{ye2016sc}. Since $\setS' \subsetsim \setS$ we expect $\theta_i \approx 0$ and hence $\mC \approx \mtx{I}$, and when $\setS' \subseteq \setS$ exactly then $\mC = \mtx{I}$. Partition $\mW$ into $\dim(\setS) \times \dim(\setS')$ and $\dim(\setS) \times \dim(\setS) - \dim(\setS')$ submatrices $\mW_C$ and $\mW_I$ such that 
\begin{align*}
   \mV_{\setS} (\mtx{I} - \mG) \mV_{\setS}^H =  \mV_{\setS} \mW_C(\mtx{I} - \mC) \mW_C^H\mV_{\setS}^H + \mV_{\setS} \mW_I\mW_I^H\mV_{\setS}^H.
\end{align*}
Define the $N \times \dim(\setS) - \dim(\setS')$ matrix $\mL = \mV_{\setS} \mW_I$, and we have
\begin{align*}
    \vy - \mP_{\setS'}\vy &=\underbrace{\mL\mL^H}_{\text{low-rank}}\vy + \underbrace{\mV_{\setS} \mW_C(\mtx{I} - \mC) \mW_C^H\mV_{\setS}^H}_{\|\cdot\|~\text{small}}\vy -\underbrace{\mP_{\setS^{\perp}}\mP_{\setS'}\mP_{\setS}}_{\|\cdot\|~\text{small}}\vy.
\end{align*}
Therefore is reasonable to approximate the projection of $\vy$ onto $\setS'$ as 
\begin{align*}
    \mP_{\setS'}\vy \approx \vy - \mL\mL^H\vy,
\end{align*}
which only requires $2\cdot N \cdot (\dim(\setS) - \dim(\setS'))$ computations. The number of computations scale with the difference in the subspace dimensions as opposed to the subspace dimensions themselves. Nearly nothing is lost empirically by merging packets via a low rank update (see Table~\ref{tab:lr_update_error}).

\begin{table}[t]
    \centering
    \begin{tabular}{|c|c|c|c|c|}
        \hline
         Subspace Dim. & $\lceil 2\Omega T_{KN}(\theta)\rceil + 2$  & $\lceil 2\Omega T_{KN}(\theta)\rceil + 4$ & $\lceil 2\Omega T_{KN}(\theta)\rceil + 6$ & $\lceil 2\Omega T_{KN}(\theta)\rceil + 8$  \\
         \hline
           $\lceil 2\Omega T_{N}(\theta)\rceil + 2$ &  $2.8 \times 10^{-4}$   &  $2.0 \times 10^{-3}$ &  $3.9 \times 10^{-3}$  &  $1.3 \times 10^{-2}$ \\
          \hline
          $\lceil 2\Omega T_{N}(\theta)\rceil + 4$ &  $1.7 \times 10^{-4}$ &  $1.4 \times 10^{-3}$ &  $3.4 \times 10^{-3}$ &  $8.4 \times 10^{-3}$  \\
          \hline
          $\lceil 2\Omega T_{N}(\theta)\rceil + 6$ &  $1.16 \times 10^{-4}$ &  $9.6 \times 10^{-4}$ &  $1.5 \times 10^{-3}$ &  $5.4 \times 10^{-3}$   \\
          \hline
          $\lceil 2\Omega T_{N}(\theta)\rceil + 8$ &  $5.9 \times 10^{-5}$ &  $3.42 \times 10^{-4}$ &  $1 \times 10^{-3}$ &  $3.1 \times 10^{-3}$  \\
          \hline
          $\lceil 2\Omega T_{N}(\theta)\rceil + 10$ &  $3.1 \times 10^{-5}$ &  $1.5 \times 10^{-4}$ &  $2.6 \times 10^{-4}$ &  $1 \times 10^{-3}$  \\
          \hline
    \end{tabular}
    \caption{\small \sl A comparison of $\| \mP_{\setS'}\mP_{\setS} - (\mP_{\setS}-\mL\mL^H\mP_{\setS})\|_F/\| \mP_{\setS'}\mP_{\setS}\|_F$, e.g.\ a measure of the low-rank updates accuracy, to several choices in packet subspace dimension (columns) and merged packet subspace dimension (rows) for a ULA. The single packet batch size is $N=2^6$, and $\lceil 2\Omega T_{N}(\theta)\rceil = 79$. We merge 5 packets so that $\lceil 2\Omega T_{KN}(\theta)\rceil = 335$, and the rank of the low rank update is set to $\dim(\setS) - \dim(\setS')$.}
    \label{tab:lr_update_error}
\end{table}

To test the validity of our claim that a low rank update can be used as an approximation to a direct projection onto $\setS'$ we return to the ULA experiment used in Section~\ref{sec:streaming}. We again use packets with a batch size of $N=2^6$, and derive the low rank update for the case where we merge 5 packets. We use $\| \mP_{\setS'}\mP_{\setS} - (\mP_{\setS}-\mL\mL^H\mP_{\setS})\|_F/\| \mP_{\setS'}\mP_{\setS}\|_F$ as our metric, which compares how ``close" the low rank approximate is to the composite projection $\mP_{\setS'}\mP_{\setS}$. The subspace dimension on the packet and merged packet level were varied to see how performance is affected. The results shown in Table~\ref{tab:lr_update_error} indicate that as we increase the packet subspace dimension the the approximation quality increases. This is likely due to the misalignment term being reduced as the merged packet subspace becomes closer to being contained in the direct union of packet subspaces. Therefore it may be beneficial to overestimate the subspace dimension at the single packet level prior to combining.

\end{document}